\errorstopmode
\input amssym.def
\input amssym.tex


\magnification=\magstephalf
\hsize=14.0 true cm
\vsize=19 true cm
\hoffset=1.0 true cm
\voffset=2.0 true cm

\abovedisplayskip=12pt plus 3pt minus 3pt
\belowdisplayskip=12pt plus 3pt minus 3pt
\parindent=1.0em


\font\sixrm=cmr6
\font\eightrm=cmr8
\font\ninerm=cmr9

\font\sixi=cmmi6
\font\eighti=cmmi8
\font\ninei=cmmi9

\font\sixsy=cmsy6
\font\eightsy=cmsy8
\font\ninesy=cmsy9

\font\sixbf=cmbx6
\font\eightbf=cmbx8
\font\ninebf=cmbx9

\font\eightit=cmti8
\font\nineit=cmti9

\font\eightsl=cmsl8
\font\ninesl=cmsl9

\font\sixss=cmss8 at 8 true pt
\font\sevenss=cmss9 at 9 true pt
\font\eightss=cmss8
\font\niness=cmss9
\font\tenss=cmss10

\font\eighttt=cmtt8
\font\ninett=cmtt9
\font\tentt=cmtt10

\font\fivemib=cmmib5
\font\sixmib=cmmib6
\font\sevenmib=cmmib7
\font\eightmib=cmmib8
\font\ninemib=cmmib9
\font\tenmib=cmmib10

\font\fivesyb=cmbsy5
\font\sevensyb=cmbsy7
\font\tensyb=cmbsy10

 at 12 true pt
 at 12 true pt
\font\bigrm=cmr10 at 12 true pt
 at 12 true pt
 at 12 true pt

 at 16 true pt
 at 16 true pt
 at 16 true pt
 at 16 true pt
 at 16 true pt

\catcode`@=11
\newfam\mibfam
\newfam\ssfam

\def\tenpoint{\def\rm{\fam0\tenrm}%
    \textfont0=\tenrm \scriptfont0=\sevenrm \scriptscriptfont0=\fiverm
    \textfont1=\teni  \scriptfont1=\seveni  \scriptscriptfont1=\fivei
    \textfont2=\tensy \scriptfont2=\sevensy \scriptscriptfont2=\fivesy
    \textfont3=\tenex \scriptfont3=\tenex   \scriptscriptfont3=\tenex
    \textfont\itfam=\tenit                  \def\it{\fam\itfam\tenit}%
    \textfont\slfam=\tensl                  \def\sl{\fam\slfam\tensl}%
    \textfont\bffam=\tenbf \scriptfont\bffam=\sevenbf
                           \scriptscriptfont\bffam=\fivebf
                           \def\bf{\fam\bffam\tenbf}%
    \textfont\ssfam=\tenss \scriptfont\ssfam=\sevenss
                           \scriptscriptfont\ssfam=\sevenss
                           \def\ss{\fam\ssfam\tenss}%
    \textfont\ttfam=\tentt \def\tt{\fam\ttfam\tentt}%
    \textfont\mibfam=\tenmib \scriptfont\mibfam=\sevenmib
                             \scriptscriptfont\mibfam=\sevenmib
                             \def\mib{\fam\mibfam\tenmib}%
    \normalbaselineskip=13pt
    \setbox\strutbox=\hbox{\vrule height8.5pt depth3.5pt width0pt}%
    \let\big=\tenbig
    \normalbaselines\rm}

\def\ninepoint{\def\rm{\fam0\ninerm}%
    \textfont0=\ninerm      \scriptfont0=\sixrm
                            \scriptscriptfont0=\fiverm
    \textfont1=\ninei       \scriptfont1=\sixi
                            \scriptscriptfont1=\fivei
    \textfont2=\ninesy      \scriptfont2=\sixsy
                            \scriptscriptfont2=\fivesy
    \textfont3=\tenex       \scriptfont3=\tenex
                            \scriptscriptfont3=\tenex
    \textfont\itfam=\nineit \def\it{\fam\itfam\nineit}%
    \textfont\slfam=\ninesl \def\sl{\fam\slfam\ninesl}%
    \textfont\bffam=\ninebf \scriptfont\bffam=\sixbf
                            \scriptscriptfont\bffam=\fivebf
                            \def\bf{\fam\bffam\ninebf}%
    \textfont\ssfam=\niness \scriptfont\ssfam=\sixss
                            \scriptscriptfont\ssfam=\sixss
                            \def\ss{\fam\ssfam\niness}%
    \textfont\ttfam=\ninett \def\tt{\fam\ttfam\ninett}%
    \textfont\mibfam=\ninemib \scriptfont\mibfam=\sixmib
                            \scriptscriptfont\mibfam=\sixmib
                            \def\mib{\fam\mibfam\ninemib}%
    \normalbaselineskip=12pt
    \setbox\strutbox=\hbox{\vrule height8.0pt depth3.0pt width0pt}%
    \let\big=\ninebig
    \normalbaselines\rm}

\def\eightpoint{\def\rm{\fam0\eightrm}%
    \textfont0=\eightrm      \scriptfont0=\sixrm
                             \scriptscriptfont0=\fiverm
    \textfont1=\eighti       \scriptfont1=\sixi
                             \scriptscriptfont1=\fivei
    \textfont2=\eightsy      \scriptfont2=\sixsy
                             \scriptscriptfont2=\fivesy
    \textfont3=\tenex        \scriptfont3=\tenex
                             \scriptscriptfont3=\tenex
    \textfont\itfam=\eightit \def\it{\fam\itfam\eightit}%
    \textfont\slfam=\eightsl \def\sl{\fam\slfam\eightsl}%
    \textfont\bffam=\eightbf \scriptfont\bffam=\sixbf
                             \scriptscriptfont\bffam=\fivebf
                             \def\bf{\fam\bffam\eightbf}%
    \textfont\ssfam=\eightss \scriptfont\ssfam=\sixss
                             \scriptscriptfont\ssfam=\sixss
                             \def\ss{\fam\ssfam\eightss}%
    \textfont\ttfam=\eighttt \def\tt{\fam\ttfam\eighttt}%
    \textfont\mibfam=\eightmib \scriptfont\mibfam=\sixmib
                             \scriptscriptfont\mibfam=\sixmib
                             \def\mib{\fam\mibfam\eightmib}%
    \normalbaselineskip=10pt
    \setbox\strutbox=\hbox{\vrule height7.0pt depth2.0pt width0pt}%
    \let\big=\eightbig
    \normalbaselines\rm}

\def\tenbig#1{{\hbox{$\left#1\vbox to8.5pt{}\right.\n@space$}}}
\def\ninebig#1{{\hbox{$\textfont0=\tenrm\textfont2=\tensy
                       \left#1\vbox to7.25pt{}\right.\n@space$}}}
\def\eightbig#1{{\hbox{$\textfont0=\ninerm\textfont2=\ninesy
                       \left#1\vbox to6.5pt{}\right.\n@space$}}}

\def\setmathbold{
\textfont1=\tenmib \scriptfont1=\sevenmib \scriptscriptfont1=\fivemib
\textfont2=\tensyb \scriptfont2=\sevensyb \scriptscriptfont2=\fivesyb}

\def\unsetmathbold{
\textfont1=\teni \scriptfont1=\seveni \scriptscriptfont1=\fivei
\textfont2=\tensy \scriptfont2=\sevensy \scriptscriptfont2=\fivesy
}

\font\sectionfont=cmbx10
\font\subsectionfont=cmti10

\def\figurecaptionfont{\ninepoint}
\def\tablecaptionfont{\ninepoint}
\def\footnotefont{\eightpoint}


\newcount\equationno
\newcount\bibitemno
\newcount\figureno
\newcount\tableno

\equationno=0
\bibitemno=0
\figureno=0
\tableno=0


\footline={\ifnum\pageno=0{\hfil}\else
{\hss\rm\the\pageno\hss}\fi}


\def\section #1. #2 \par
{\vskip0pt plus .10\vsize\penalty-100 \vskip0pt plus-.10\vsize
\vskip 1.6 true cm plus 0.2 true cm minus 0.2 true cm
\global\def\equationlabel{#1}
\global\equationno=0
\setmathbold
\leftline{\sectionfont #1. #2}\par
\immediate\write\terminal{Section #1. #2}\unsetmathbold
\vskip 0.7 true cm plus 0.1 true cm minus 0.1 true cm
\noindent}


\def\subsection #1 \par
{\vskip0pt plus 1.0 true cm\penalty-50 \vskip0pt plus-1.0 true cm
\vskip2.5ex plus 0.1ex minus 0.1ex
\leftline{\subsectionfont #1}\par
\immediate\write\terminal{Subsection #1}
\vskip1.0ex plus 0.1ex minus 0.1ex
\noindent}


\def\appendix #1. #2 \par
{\vskip0pt plus .10\vsize\penalty-100 \vskip0pt plus-.10\vsize
\vskip 1.6 true cm plus 0.2 true cm minus 0.2 true cm
\global\def\equationlabel{\hbox{\rm#1}}
\global\equationno=0
\setmathbold
\leftline{\sectionfont Appendix #1. #2}\par
\immediate\write\terminal{Appendix #1. #2}\unsetmathbold
\vskip 0.7 true cm plus 0.1 true cm minus 0.1 true cm
\noindent}



\def\equation#1{$$\displaylines{\qquad #1}$$}
\def\enum{\global\advance\equationno by 1
\hfill\llap{{\rm(\equationlabel.\the\equationno)}}}

\def\next#1{\cr\noalign{\vskip#1}\qquad}
\def\nexteq#1{\cr\noalign{\vskip#1}\qquad}


\def\ifundefined#1{\expandafter\ifx\csname#1\endcsname\relax}

\def\ref#1{\ifundefined{#1}?\immediate\write\terminal{unknown reference
on page \the\pageno}\else\csname#1\endcsname\fi}

\newwrite\terminal
\newwrite\bibitemlist

\def\bibitem#1#2\par{\global\advance\bibitemno by 1
\immediate\write\bibitemlist{\string\def
\expandafter\string\csname#1\endcsname
{\the\bibitemno}}
\item{[\the\bibitemno]}#2\par}

\def\beginbibliography{
\vskip0pt plus .15\vsize\penalty-100 \vskip0pt plus-.15\vsize
\vskip 1.2 true cm plus 0.2 true cm minus 0.2 true cm
\leftline{\sectionfont References}\par
\immediate\write\terminal{References}
\immediate\openout\bibitemlist=biblist
\frenchspacing\parindent=1.8em
\vskip 0.5 true cm plus 0.1 true cm minus 0.1 true cm}

\def\endbibliography{
\immediate\closeout\bibitemlist
\nonfrenchspacing\parindent=1.0em}


\def\figurecaption#1{\global\advance\figureno by 1
\narrower\figurecaptionfont Fig.~\the\figureno. #1}

\def\tablecaption#1{\global\advance\tableno by 1
\centerline{\tablecaptionfont Table~\the\tableno. #1}}

\def\thicktablerule{\hrule height0.8pt}
\def\thintablerule{\hrule height0.4pt}

\tenpoint

\immediate\openin\bibitemlist=biblist
\ifeof\bibitemlist\immediate\closein\bibitemlist
\else\immediate\closein\bibitemlist
\input biblist \fi


\def\thismonth{\ifcase\month\or
January\or February\or March\or April\or May\or June\or
July\or August\or September\or October\or November\or December\fi}

\font\fourteenrm=cmr12 scaled \magstep1
\font\fourteeni=cmmi12 scaled \magstep1
\font\fourteensy=cmsy10 scaled \magstep2
\font\fourteenex=cmex10 scaled \magstep2
\font\fourteenbf=cmbx12 scaled \magstep1
\font\fourteensl=cmsl12 scaled \magstep1
\font\fourteentt=cmtt12 scaled \magstep1
\font\fourteenit=cmti12 scaled \magstep1

\ifx\fourteenpoint\undefined
   \def\fourteenpoint{\def\rm{\fam0\fourteenrm}
       \textfont0=\fourteenrm \scriptfont0=\tenrm \scriptscriptfont0=\sevenrm
       \textfont1=\fourteeni  \scriptfont1=\teni  \scriptscriptfont1=\seveni
       \textfont2=\fourteensy \scriptfont2=\tensy \scriptscriptfont2=\sevensy
       \textfont3=\fourteenex \scriptfont3=\fourteenex
                              \scriptscriptfont3=\fourteenex
       \textfont\itfam=\fourteenit  \def\it{\fam\itfam\fourteenit}%
       \textfont\slfam=\fourteensl  \def\sl{\fam\slfam\fourteensl}%
       \textfont\ttfam=\fourteentt  \def\tt{\fam\ttfam\fourteentt}%
       \textfont\bffam=\fourteenbf  \scriptfont\bffam=\tenbf
        \scriptscriptfont\bffam=\sevenbf  \def\bf{\fam\bffam\fourteenbf}%
       \normalbaselineskip=17pt
       \setbox\strutbox=\hbox{\vrule height11.9pt depth6.3pt width0pt}%
       \normalbaselines\rm}
   \fi

\font\seventeenrm=cmr17
\font\seventeeni=cmmi12 scaled \magstep2
\font\seventeensy=cmsy10 scaled \magstep3
\font\seventeenex=cmex10 scaled \magstep3
\font\seventeenbf=cmbx12 scaled \magstep2
\font\seventeensl=cmsl12 scaled \magstep2
\font\seventeentt=cmtt12 scaled \magstep2
\font\seventeenit=cmti12 scaled \magstep2

\font\twelverm=cmr12
\font\twelvei=cmmi12
\font\twelvesy=cmsy10 scaled \magstep1
\font\twelvebf=cmbx12

\ifx\seventeenpoint\undefined
   \def\seventeenpoint{\def\rm{\fam0\seventeenrm}
       \textfont0=\seventeenrm \scriptfont0=\twelverm \scriptscriptfont0=\ninerm
       \textfont1=\seventeeni  \scriptfont1=\twelvei  \scriptscriptfont1=\ninei
       \textfont2=\seventeensy \scriptfont2=\twelvesy \scriptscriptfont2=\ninesy
       \textfont3=\seventeenex \scriptfont3=\seventeenex
                              \scriptscriptfont3=\seventeenex
       \textfont\itfam=\seventeenit  \def\it{\fam\itfam\seventeenit}%
       \textfont\slfam=\seventeensl  \def\sl{\fam\slfam\seventeensl}%
       \textfont\ttfam=\seventeentt  \def\tt{\fam\ttfam\seventeentt}%
       \textfont\bffam=\seventeenbf  \scriptfont\bffam=\twelvebf
        \scriptscriptfont\bffam=\ninebf  \def\bf{\fam\bffam\seventeenbf}%
       \normalbaselineskip=21pt
       \setbox\strutbox=\hbox{\vrule height17pt depth4pt width0pt}%
       \normalbaselines\rm}
   \fi

\input epsf
\epsfclipon



\def\rmd{{\rm d}}

\def\rme{{\rm e}}
\def\rmO{{\rm O}}

\def\urltilde{\kern -.15em\lower .7ex\hbox{\~{}}\kern .04em}



\def\proof{\noindent{\sl Proof:}\kern0.6em}

\def\frac#1#2{\hbox{$#1\over#2$}}
\def\dual{\mathstrut^*\kern-0.1em}

\def\ring{\mathaccent"7017}
\def\lvec#1{\setbox0=\hbox{$#1$}
    \setbox1=\hbox{$\scriptstyle\leftarrow$}
    #1\kern-\wd0\smash{
    \raise\ht0\hbox{$\raise1pt\hbox{$\scriptstyle\leftarrow$}$}}
    \kern-\wd1\kern\wd0}
\def\rvec#1{\setbox0=\hbox{$#1$}
    \setbox1=\hbox{$\scriptstyle\rightarrow$}
    #1\kern-\wd0\smash{
    \raise\ht0\hbox{$\raise1pt\hbox{$\scriptstyle\rightarrow$}$}}
    \kern-\wd1\kern\wd0}
\def\cvec#1{\kern-0.5pt\vec{\kern0.5pt #1}}

\def\slash#1{\setbox2=\hbox{$\displaystyle#1$}%
             \setbox3=\hbox{$\displaystyle/$}%
             #1\kern-0.8\wd2/\kern-1.0\wd3\kern0.8\wd2\kern0.5pt}

\def\wick#1{\setbox2=\hbox{$\displaystyle#1$}
    \setbox3=\null\ht3=3.0pt\dp3=0.0pt\wd3=20.0pt
    #1\kern-\wd2\kern3.0pt\raise11.0pt\vbox{\hrule height0.3pt
    \hbox{\vrule width0.3pt\box3\vrule width0.3pt}}\kern-24.0pt\kern\wd2}

\def\longwick#1{\setbox2=\hbox{$\displaystyle#1$}
    \setbox3=\null\ht3=3.0pt\dp3=0.0pt\wd3=27.0pt
    #1\kern-\wd2\kern3.0pt\raise11.0pt\vbox{\hrule height0.3pt
    \hbox{\vrule width0.3pt\box3\vrule width0.3pt}}\kern-31.0pt\kern\wd2}

\def\verylongwick#1{\setbox2=\hbox{$\displaystyle#1$}
    \setbox3=\null\ht3=3.0pt\dp3=0.0pt\wd3=43.0pt
    #1\kern-\wd2\kern3.0pt\raise11.0pt\vbox{\hrule height0.3pt
    \hbox{\vrule width0.3pt\box3\vrule width0.3pt}}\kern-47.0pt\kern\wd2}


\def\nab#1{{\nabla_{#1}}}
\def\nabstar#1{{\nabla\kern0.5pt\smash{\raise 4.5pt\hbox{$\ast$}}
               \kern-5.5pt_{#1}}}
\def\drv#1{{\partial_{#1}}}
\def\drvstar#1{{\partial\kern0.5pt\smash{\raise 4.5pt\hbox{$\ast$}}
               \kern-6.0pt_{#1}}}
\def\sdrvstar#1{{\partial\kern0.4pt\smash{\raise 3.6pt\hbox{$\ast$}}
                \kern-4.8pt_{#1}}}

\def\ldrvstar#1{{\lvec{\,\partial}\kern-0.5pt\smash{\raise 4.5pt\hbox{$\ast$}}
               \kern-5.0pt_{#1}}}


\def\MeV{{\rm MeV}}

\def\fm{{\rm fm}}
\def\MSbar{\overline{\rm MS\kern-0.5pt}\kern0.5pt}



\def\imp#1{#1_{\hbox{\sixrm I}}}


\def\dirac#1{\gamma_{#1}}
\def\diracstar#1#2{
    \setbox0=\hbox{$\gamma$}\setbox1=\hbox{$\gamma_{#1}$}
    \gamma_{#1}\kern-\wd1\kern\wd0
    \smash{\raise4.5pt\hbox{$\scriptstyle#2$}}}


\def\SUthree{{\rm SU(3)}}

\def\tr{{\rm tr}}


\def\SG{S_{\rm G}}
\def\Spf{S_{\rm pf}}


\def\csw{c_{\rm sw}}
\def\cA{c_{\rm A}}
\def\cF{c_{\rm F}}
\def\Dee{D_{\rm ee}}
\def\Deo{D_{\rm eo}}
\def\Doe{D_{\rm oe}}
\def\Doo{D_{\rm oo}}
\def\Dhat{\hat{D}}


\def\Ai{${\rm A}_1$}
\def\Aii{${\rm A}_2$}
\def\Aiii{${\rm A}_3$}
\def\Bi{${\rm B}_1$}
\def\Ci{${\rm X}_1$}


\def\Fpi{F_{\pi}}

\def\FK{F_{K}}

\def\Mpi{M_{\pi}}
\def\MK{M_K}
\def\ZA{Z_A}
\def\ZAh{\hat{Z}_A}


\def\evm{\bar{\mu}}
\def\evs{\sigma}


\def\eps{\epsilon}
\def\Pacc{P_{\rm acc}}

\def\rspike{R_{\rm spk}}

\def\xvec{\cvec{x}}

\def\ha{\tilde{h}}
\def\st{\tilde{\sigma}}
\def\rt{\tilde{\rho}}
\def\Ft{\tilde{F}}
\def\erf{{\rm erf}}
\def\nrm#1{\|#1\|_2}
\def\inrm#1{\|#1\|_{\infty}}
\def\DD{D^{\dagger}\kern-1pt D}
\def\pbar{\kern0.5pt\overline{\kern-0.5pt w\kern-0.5pt}_p\kern0.5pt}

%
\rightline{CERN-TH-2019-182}
\vskip1.2cm
{\fourteenpoint
\centerline{Master-field simulations of O($a$)-improved lattice QCD:}
\vskip 0.2 true cm
\centerline{Algorithms, stability and exactness}
}

\tenpoint
\vskip 0.6 true cm
\centerline{\bigrm Anthony Francis$^{\rm a}$, Patrick Fritzsch$^{\rm a}$,
Martin L\"uscher$^{\rm a,b}$ and Antonio Rago$^{\rm c}$}

\vskip1.5ex
\centerline{{\it $^{\rm a}$CERN,
Theoretical Physics Department, 1211 Geneva 23, Switzerland}}

\vskip1.0ex
\centerline{{\it $^{\rm b}$Albert Einstein Center for Fundamental Physics}}
\centerline{{\it Institute for Theoretical Physics,
Sidlerstrasse 5, 3012 Bern, Switzerland}}

\vskip1.0ex
\centerline{{\it $^{\rm c}$Centre for Mathematical Sciences,
University of Plymouth}}
\centerline{{\it Plymouth, PL4 8AA, United Kingdom}}

\vskip 0.8 true cm
\thintablerule
\vskip 2.0ex
\ninepoint
\noindent
In master-field simulations of lattice QCD, the expectation values
of interest are obtained from a single or at most a few
representative gauge-field configurations on very large lattices.
If the light quarks are included,
the generation of these fields using standard
techniques is however challenging
in view of various algorithmic instabilities and precision issues.
Ways to overcome these problems are described here
for the case of the O($a$)-improved Wilson formulation
of lattice QCD
and the viability of the proposed measures is then checked
in extensive simulations of the theory
with $2+1$ flavours of quarks.

\vskip 2.0ex
\thintablerule

\tenpoint


\section 1. Introduction

Numerical lattice QCD usually proceeds by generating a representative
ensemble of gauge-field configurations through a Markov process and
estimating the expectation values of the chosen observables
through ensemble averages.
On very large lattices, translation averages
in presence of a single gauge field (the master field)
provide an alternative way of calculating the expectation values,
the associated statistical errors being
determined through translation averages too [\ref{LuscherMaster}].

In the case of the SU(3) gauge theory in four dimensions,
some large-scale simulations of this kind were recently
performed and worked out as expected [\ref{HighTop}].
While the inclusion of the light quarks in these calculations is
in principle straightforward, some technical questions
must be answered before the established lattice QCD
techniques can be applied with confidence. Global accept-reject
steps, for example, can be a source of inexactness as a result of
significance losses growing proportionally to the lattice size.
Other possible issues include numerical instabilities of the
simulation algorithm and unbalanced
local inaccuracies of approximately calculated
solutions of the Dirac equation.

In the present paper, all these potential obstacles for
master-field simulations are addressed and
solutions are brought forward
in each case. For definiteness Wilson's formulation of lattice QCD
is considered [\ref{Wilson}], with O($a$) counterterms added as usual
[\ref{SW},\ref{SFimp}], but the material covered in sects.~3--5 is
not specific to this form of the lattice theory.
Moreover, an enhanced stability of the simulations
is expected to be be\-ne\-ficial for traditional QCD simulations too.

The proposed stabilizing measures include
a slight modification of the standard O($a$)-improved
lattice Dirac operator (sect.~2) and the use of
the Stochastic Molecular Dynamics (SMD) simulation algorithm
[\ref{Horowitz},\ref{JansenLiu}]
in place of the Hybrid Monte Carlo (HMC) algorithm [\ref{HMC}] (sect.~3).
Which level of numerical precision is required for the algorithm
to simulate the theory exactly,
and how sufficient precision can be guaranteed,
is discussed in sects.~4 and 5.
The results of some representative
simulations of QCD with $2+1$ flavours of quarks,
demonstrating the viability of the proposed framework,
are then reported in sect.~6.

\section 2. O($a$)-improvement revisited

Accidental near-zero modes of
the Wilson--Dirac operator are commonly suspected to
cause simulation instabilities and enhanced statistical
fluctuations.
Experience moreover suggests that the addition of the Pauli term required
for O($a$)-improvement tends to further these
undesirable effects, particularly
so on coarse lattices. In this section,
a modified improved Dirac operator is introduced,
which can be expected to be better behaved in this respect
and which has some other advantages too.

\subsection 2.1 Preliminaries

The lattice theory is set up on four-dimensional hypercubic
lattices with time extent $T$, spatial size $L^3$, spacing $a$ and
periodic boundary conditions in the space directions.
Either periodic (anti-periodic for the quark fields),
Schr\"odinger-functional [\ref{SF},\ref{SFQ}] or
open boundary conditions [\ref{OpenBC},\ref{TMopenQCD}]
are imposed in the time direction.
The gauge group is taken to be SU(3) and
the notational conventions for the group generators, the
link variables, etc., are as in
refs.~[\ref{SFimp},\ref{OpenBC},\ref{TMopenQCD}].

In the simulations reported in sect.~6,
the tree-level Symanzik-improved gauge action is used
[\ref{TreeSymanzik}], but the choice of the gauge action is
otherwise unimportant.
For notational convenience, lattice units are usually employed,
where all dimensionful quantities are expressed in units of
the lattice spacing $a$.

\subsection 2.2 Lattice Dirac operator

On a lattice with periodic boundary conditions, the O($a$)-improved
Wilson--Dirac operator is given by
\equation{
  D=\frac{1}{2}\left\{
  \dirac{\mu}(\nabstar{\mu}+\nab{\mu})-\nabstar{\mu}\nab{\mu}\right\}
  +\csw\frac{i}{4}
  \sigma_{\mu\nu}\widehat{F}_{\mu\nu}+m_0,
  \enum
}
where $m_0$ is the bare quark mass, $\nab{\mu}$ and
$\nabstar{\mu}$ the forward and backward gauge-covariant
difference operators and $\widehat{F}_{\mu\nu}$ the
standard (``clover'') lattice expression [\ref{SFimp}]
for the gauge-field tensor. The coefficient $\csw$
is equal to $1$ at tree-level of perturbation theory and
grows monotonically with the gauge coupling, typically
reaching values around $2$ on coarse lattices
(see ref.~[\ref{CswThree}], for example).

If the lattice points are classified as even or odd depending on
the parity of the sum of their coordinates, the Dirac operator
assumes the block form
\equation{
  D=\pmatrix{\Dee & \Deo \cr
             \noalign{\vskip1.0ex}
             \Doe & \Doo \cr},
  \enum
}
$\Deo$ and $\Doe$ being the hopping terms from the odd to the even
points and the even to the odd points, respectively, while
the Pauli term is included in the diagonal
part
\equation{
  \Dee+\Doo=M_0+
  \csw\frac{i}{4}
  \sigma_{\mu\nu}\widehat{F}_{\mu\nu},
  \qquad M_0=4+m_0.
  \enum
}
A simple way to accelerate lattice QCD simulations exploits this
structure by passing to the even-odd preconditioned form
$\Dhat=\Dee-\Deo\Doo^{-1}\Doe$
of the lattice Dirac operator, which acts on quark spinors on the
even lattice sites only.

The Pauli term in these equations can be
fairly large, particularly so on coarse lattices,
where it may get close to saturating the norm bound
\equation{
  \bigl\|\frac{i}{4}
  \sigma_{\mu\nu}\widehat{F}_{\mu\nu}\bigr\|_2\leq 3.
  \enum
}
Since the positive and negative
eigenvalues of the Pauli term are equally distributed, and
the bare mass $m_0$ is usually negative, the diagonal
part of the Dirac operator is then not protected
from having arbitrarily small eigenvalues.
Even-odd preconditioning is in fact known to occasionally fail
for this reason, with probability growing proportionally
to the lattice size, which practically excludes its
use in master-field simulations.

\subsection 2.3 Alternative form of the lattice Dirac operator

The fact that the diagonal part of the improved Wilson--Dirac operator
is not positive distinguishes the improved from the unimproved operator
and could explain why the improvement tends to promote the
instabilities mentioned at the beginning of this section.

From the point of view of
O($a$)-improvement and the continuum limit,
the alternative expression
\equation{
  \Dee+\Doo=M_0
  \exp\biggl\{{\csw\over4+m_0}\frac{i}{4}
  \sigma_{\mu\nu}\widehat{F}_{\mu\nu}\biggr\}
  \enum
}
for the diagonal part of the Dirac operator
may however do just as well.
At leading order of perturbation theory,
this expression actually
coincides with the traditional one and improvement is
achieved by setting $\csw=1$.
Clearly, this form of the diagonal part of the Dirac operator
is positive definite and safely invertible.
Even-odd preconditioning is therefore guaranteed to
be numerically unproblematic.
Moreover, $\det D=\det\Dhat$ up to a field-independent
proportionality constant.

Whether the alternative form (2.5)
of the diagonal part of the Dirac operator
is a viable choice at all couplings in the
scaling regime is an open question at this point, which
must ultimately be answered through extensive simulations
of the modified theory.
In these simulations,
the exponential of the Pauli term
and the derivative of the exponential with respect to the gauge field
must be frequently evaluated,
but there are ways to do this
with negligible computational effort (see appendix A).

\subsection 2.4 Including O($a$) boundary counterterms

If open or Schr\"odinger-functional boundary
conditions are imposed in the time direction, O($a$)-improvement
near the lattice boundaries requires
boundary counterterms to be added to the quark
action. The counterterms amount to
replacing the constant $M_0$ in eq.~(2.3) by a
diagonal operator that acts on quark fields $\psi(x)$
according to [\ref{SFimp},\ref{OpenBC}]
\equation{
   M_0\psi(x)=\left\{4+m_0+(\cF-1)(\delta_{x_0,1}+\delta_{x_0,T-1})
   \right\}\psi(x).
   \enum
}
At tree-level of perturbation theory,
the improvement coefficient $\cF$ is equal to $1$ and the
boundary term in eq.~(2.6) thus vanishes at this order.

The alternative expression for the diagonal part of the improved
Dirac operator is then again given by eq.~(2.5) with $M_0$
set to the operator (2.6). This way of adding the
boundary counterterms preserves all the good properties
of the term mentioned above.

\section 3. Stochastic molecular dynamics

The numerical integration of the molecular-dynamics equations
in the HMC algorithm is known to occasionally go astray,
leading to unbounded violations of energy conservation.
In these singular cases, the time-reversibility of the integration
is likely to be violated too and the exactness of
the simulation is then no longer guaranteed.
What causes these instabilities remains unclear.
In particular,
they occur even if the lattice Dirac operator is rigorously
protected from having near-zero modes
(through a twisted-mass term, for example).

With respect to the HMC algorithm,
the Stochastic Molecular Dynamics (SMD) simulation algorithm
[\ref{Horowitz},\ref{JansenLiu}]
described in this section tends to be less affected
by integration instabilities.
One of the reasons for this
favourable behaviour is the typically much shorter
molecular-dynamics integration time, but the fact that
the pseudo-fermion fields are more frequently adapted
to the changes of the gauge field may have a stabilizing
effect too.

\subsection 3.1 SMD update cycle

The SMD algorithm is rather similar to the HMC algorithm and like
the latter does not assume a particular form of the lattice action.
For simplicity the algorithm is here
described for two-flavour QCD and no frequency-splitting of
the quark determinant, the generalization
to other cases of interest being straightforward.

The fields processed by the SMD algorithm are the gauge field
$U(x,\mu)$, the associated momentum field $\pi(x,\mu)$ and
a pseudo-fermion field $\phi(x)$ with action
\equation{
  \Spf(U,\phi)=(\phi,(\DD)^{-1}\phi).
  \enum
}
In the case considered, the only other term included in the total action
$S(U,\phi)$ of the theory is the gauge action $\SG(U)$.
An SMD update cycle then consists of a random rotation of the
momentum and the pseudo-fermion field, a short
molecular-dynamics evolution of the momentum and the gauge field
and, finally, an accept-reject step that makes the algorithm exact.

\subsection 3.2 Random field rotation

At the beginning of the update cycle,
the momentum and the pseudo-fermion field are refreshed according to
\equation{
  \pi\to c_1\pi+c_2\upsilon,
  \qquad
  c_1=\rme^{-\gamma\eps},\quad c_2=(1-c_1^2)^{1/2},
  \enum
  \nexteq{2.5ex}
  \phi\to c_1\phi+c_2D^{\dagger}\eta,
  \enum
}
where the fields $\upsilon(x,\mu)$ and $\eta(x)$
are chosen randomly with normal distribution.
Both $\gamma>0$ and $\eps>0$ are fixed parameters of the SMD algorithm,
$\eps$ being equal to the molecular-dynamics integration time
(see subsect.~3.3).

The algorithm defined by eqs.~(3.2),(3.3) simulates the Gaussian distribution
\equation{
  {\rm constant}\times\exp\{-\frac{1}{2}(\pi,\pi)-\Spf(U,\phi)\}
  \enum
}
at fixed gauge field.
If there are
further pseudo-fermion fields,
this property must be maintained
by including them in
the update step and by adapting eq.~(3.3)
to the chosen pseudo-fermion actions.
The ``friction parameter'' $\gamma$ determines how quickly
the memory of previous field configurations is lost.
In principle each field may have its own friction parameter,
but for simplicity $\gamma$ is here taken to be same for all fields.

\subsection 3.3 Molecular-dynamics evolution and accept-reject step

In the second step of the SMD update cycle, the molecular-dynamics
equations for the momentum and the gauge field, which
derive from the Hamilton function
\equation{
  H(\pi,U)=\frac{1}{2}(\pi,\pi)+S(U,\phi),
  \enum
}
are integrated from the current simulation time $t$ to
$t+\eps$ using a reversible symplectic integration rule.
The fields $\tilde{\pi}$ and $\tilde{U}$
obtained at the end of the integration are then accepted with
probability
\equation{
  \Pacc(\pi,U)=\min\{1,\rme^{-\Delta H(\pi,U)}\},
  \qquad
  \Delta H(\pi,U)=H(\tilde{\pi},\tilde{U})-H(\pi,U).
  \enum
}
Otherwise, i.e.~if the proposed fields $\tilde{\pi},\tilde{U}$
are not accepted,
the gauge field is set to $U$ and the momentum field to $-\pi$
[\ref{JansenLiu}]. Throughout this step, the pseudo-fermion field
is held fixed.

The accept-reject step guarantees that the update cycle
preserves the distribution $\rme^{-H}$ of
the gauge field, the momentum field and the pseudo-fermion field.
Moreover, if $\eps$ is sufficiently small,
the SMD algorithm can be rigorously shown to be ergodic
and to asymptotically simulate this distribution, independently of the
initial values of the fields [\ref{Ergodicity}].

\subsection 3.4 Continuous simulation-time limit

At fixed $\eps$ and large $\gamma$, the SMD algorithm
coincides with the HMC algorithm.
On the other hand, if $\eps$ is taken to zero while $\gamma$ is held fixed,
the acceptance probability (3.5) goes to $1$
and the algorithm solves the stochastic
molecular-dynamics equations
\equation{
  \partial_t\pi=-\gamma\pi-\partial_US+\upsilon,
  \enum
  \nexteq{2.0ex}
  \partial_t U=\pi U,
  \enum
  \nexteq{2.0ex}
  \partial_t\phi=-\gamma\phi+D^{\dagger}\eta.
  \enum
}
A compact notation is used here and the white-noise fields
$\upsilon$ and $\eta$ have been scaled by the factor $\sqrt{\eps/2\gamma}$
before going to the limit $\eps=0$.

The parameter $\gamma$ can in principle be set to any positive value.
Previous empirical studies of the SU(3) gauge theory
however showed that the autocorrelation
times of physical quantities have a shallow minimum around $\gamma=0.3$
[\ref{OpenBC}].
At this value of $\gamma$,
the memory of old momentum and pseudo-fermion
fields
lasts for a few molecular-dynamics time units, i.e.~for about
as long as the trajectory lengths typically chosen in HMC simulations
of lattice QCD.

\subsection 3.5 Choice of parameters and efficiency considerations

{\it SMD step size}.
Short molecular-dynamics integration times $\eps$ are beneficial for the
stability of the algorithm and do not have a negative impact on
the autocorrelation times if $\gamma$ is held fixed.
In practice a few steps of a
reversible symplectic integration rule are
applied to integrate the fields from time $t$
to $t+\eps$ and the acceptance rate $\langle\Pacc\rangle$
is set to the desired value by adjusting $\eps$.

\vskip1ex\noindent
{\it Acceptance rate}.
When field configurations are rejected, the momentum is reversed
and the algorithm tends to backtrack its trajectory in field space.
An efficient samp\-ling thus requires
the acceptance rate
to be such that rejections only occur at large separations
in simulation time, relevant reference time distances being
$1/\gamma$ and the autocorrelation times of physical quantities
[\ref{OpenBC}].

\vskip1ex\noindent
{\it Large-volume scaling}.
If all other parameters are held fixed,
the energy difference $\Delta H$ entering the acceptance
probability (3.6) grows roughly like $V^{1/2}$ with
the number $V$ of lattice points. In order to preserve
a high acceptance rate on large lattices, the integration
of the molecular-dynamics equations must thus become more and more
accurate. The computational cost per unit of simulation
time then increases proportionally to $V^{1/2p}$ if
an integration rule of order $p$ is used.
Higher-order
schemes such as the ones listed in ref.~[\ref{OMF}]
are therefore highly recommended for master-field simulations.

\vskip1ex\noindent
{\it Exploiting continuity in simulation time}.
Since the SMD algorithm updates the
fields in small steps, the computational
effort can be significantly reduced by propagating
previous solutions of the Dirac equation [\ref{chrono}]
and a local deflation subspace [\ref{DFL}]
along the trajectory in field space.
Complete regenerations of the deflation subspace are then not
even required and regular incremental updates
suffice to keep it in good condition at all times [\ref{DFL}].

\vskip1ex\noindent
{\it Simulation efficiency}.
The SMD algorithm tends to
consume more computer time per unit of
molecular-dynamics time than the HMC algorithm,
but the efficiencies of the two algorithms end up
being similar once the autocorrelation times
are taken into account [\ref{OpenBC}].

\section 4. Required numerical precision of $\Delta H$

Since the Hamilton function (3.5) is an extensive quantity,
whose average value grows proportionally to the
number of lattice points, important significance losses occur
when the energy deficits $\Delta H$ are computed at the
end of the molecular-dynamics evolution of the gauge and the
momentum field.
Depending on the lattice size and on how accurately
the Hamilton function is obtained,
the numerical errors of the calculated values of $\Delta H$
may then be such that the correctness of the simulations
is compromised by false accept-reject decisions.
The goal in this section is
to determine at which level the errors
can have a statistically noticeable
effect.

\subsection 4.1 Model distribution of $\Delta H$

On large lattices, and if the
numerical integration of the molecular-dynamics equations
is stable, the probability distribution $\rho(h)$ of $h=\Delta H$
is expected to be well represented by a Gaussian,
because $\Delta H$ is a sum of uncorrelated contributions
from distant regions on the lattice.
The distribution must satisfy
\equation{
  \int\rmd h\,\rho(h)=\int\rmd h\,\rme^{-h}\rho(h)=1
  \enum
}
and is therefore of the form
\equation{
  \rho(h)={1\over\sqrt{2\pi}\sigma}
  \exp\biggl\{-{(h-\frac{1}{2}\sigma^2)^2\over2\sigma^2}\biggr\},
  \enum
}
where the width $\sigma$ is given by the acceptance rate
\equation{
  \langle P_{\rm acc}\rangle=\int\rmd h\,\rho(h)\min\bigl\{1,\rme^{-h}\bigr\}
  =1-{\sigma\over\sqrt{2\pi}}+{\sigma^3\over24\sqrt{2\pi}}+\rmO(\sigma^5).
  \enum
}
For $\langle P_{\rm acc}\rangle=0.80$, $0.90$ and $0.95$, for example,
$\sigma$ is equal to $0.51$, $0.25$ and $0.125$.
The model distribution (4.2) is in fact accurately matched by
the empirical distributions measured in simulations like
run \Bi{} discussed in sect.~6, in which
the integration of the molecular-dynamics equations never
went astray.

The calculated approximate value $\ha$ of $\Delta H$ is distributed
slightly differently from the exact value. Assuming the numerical
errors are small, randomly distributed and uncorrelated to $h$,
the joint distribution
\equation{
  \rho_2(h,\ha)={\rho(h)\over\sqrt{2\pi}\delta}
  \exp\biggl\{{-{(\ha-h)^2\over 2\delta^2}}\biggr\}
  \enum
}
may be expected to describe the situation reasonably well,
$\delta$ being the mean square deviation of $\ha$ from $h$. Integration
over $h$ then yields the distribution
\equation{
  \rt(\ha)={1\over\sqrt{2\pi}\st}
  \exp\biggl\{-{(\ha-\frac{1}{2}\sigma^2)^2\over2\st^2}\biggr\},
  \qquad
  \st=(\sigma^2+\delta^2)^{1/2},
  \enum
}
of the actually measured values $\ha$ of $\Delta H$.

\subsection 4.2 Kolmogorov--Smirnov test

One may now ask how well the distribution $\rt$ must be sampled
to be able to distinguish it from the exact distribution $\rho$.
If a sample size of, say, $N$ trials is insufficient,
any sequence of $N$ accept-reject decisions made in an actual
simulation is statistically as likely as the one in a
(theoretical) simulation without numerical errors.

The Kolmogorov--Smirnov test provides an
answer to this question by comparing the exact cumulative
distribution
\equation{
  F(h)=\int_{-\infty}^h\rmd z\,\rho(z)
  \enum
}
with the empirical distribution
\equation{
  F_N(h)=\{\hbox{number of cases $k$ where $h_k\leq h$}\}/N
  \enum
}
obtained by
drawing $N$ numbers $h_1,\ldots,h_N$ randomly
with distribution $\rho(h)$
(see ref.~[\ref{Knuth}], for example).
In particular, the deviation
\equation{
  d=\sup_h|\Ft(h)-F(h)|
  \enum
}
of the modified cumulative distribution $\Ft$ from the exact one
can only be resolved, if $d$ is larger than
the standard deviation of $F_N$ from $F$.
Since the latter is at most $(4N)^{-1/2}$ at all values of
$h$ [\ref{Knuth}], sample sizes $N$ satisfying
\equation{
  N\geq{1\over4d^2}
  \enum
}
are required to detect the deviation (4.8) of the cumulative distributions.

\topinsert
\vbox{
\vskip0ex

\epsfxsize=7.0cm\hskip3.0cm\epsfbox{plots/ktest.eps}

\vskip2.0ex
\figurecaption{%
Number $N$ of accept-reject steps where
the difference of the distributions of the calculated
and the exact values of $\Delta H$ starts to be
statistically noticeable.
In the limit $\delta\ll\sigma$ of small numerical errors,
$N$ is asymptotically equal to $17.1\times(\sigma/\delta)^4$
(dotted line).
}
}
\endinsert

Assuming the model distributions (4.2),(4.5), some algebra shows that
\equation{
  d=\frac{1}{2}\{\erf(\st r)-\erf(\sigma r)\},
  \qquad
  r={1\over\delta}\sqrt{\ln(\st/\sigma)},
  \enum
}
is a steep function of the ratio $\delta/\sigma$.
The bound (4.9) then implies that huge statistics is
usually required before the numerical errors
of $\Delta H$ can affect the simulations at a
statistically significant level (see fig.~1).
An SMD simulation with average acceptance rate of $99\%$ and
error margin $\delta=10^{-3}$, for example,
is potentially affected only after about $7$ million
update cycles.
The precision requirements on $\Delta H$ are thus fairly mild,
a fact that was recently confirmed in
an empirical study of the SU(2) gauge theory
using the HMC algorithm [\ref{Urbach}].

\section 5. Sources of numerical inaccuracies

Lattice QCD simulations are usually performed on machines
complying with the IEEE 754 standard for floating-point
data and arithmetic. It is also common to use double-precision
(64 bit) data and operations for the basic fields, except
perhaps in some intermediate steps, when the accuracy of
the final results is provably not affected.
Various sources of numerical inaccuracies however
require special attention if large lattices
are simulated.

\subsection 5.1 Lattice sums

Scalar products of quark fields and the Hamilton function, for example,
involve a sum over all lattice points.
Important accumulations of rounding errors
in these typically huge sums can be safely avoided
using quadruple-precision (128 bit) floating-point arithmetic.
Quadruple-precision numbers may conveniently be
represented by pairs of standard IEEE 754 double-precision numbers and
there exist efficient algorithms that correctly implement
the associated arithmetic operations on any machine complying
with the standard [\ref{Dekker},\ref{Shewchuk}].

Clearly, when
the energy difference $\Delta H$ is computed, the truncation to
64 bit precision should occur only after calculating the
difference, so that $\Delta H$ is obtained with an absolute
numerical error equal to the (now practically exact) sum
of the errors of the local contributions to the Hamilton function.

\subsection 5.2 Spatially non-uniform inaccuracies

In the course of the molecular-dynamics evolution of the gauge
and the momentum field, the Dirac equation
\equation{
  D\psi(x)=\eta(x)
  \enum
}
must be solved many times for given source fields $\eta$.
The solution is obtained using some iterative algorithm, the
iteration being stopped when the current approximate
solution $\tilde{\psi}$ satisfies
\equation{
  \|\eta-D\tilde{\psi}\|\leq\omega\|\eta\|
  \enum
}
for some norm $\|\cdot\|$ and specified tolerance $\omega$.
In practice the rounding errors involved in the process set a lower limit
(about $10^{-14}$ in the case of 64 bit data and arithmetic)
on the tolerances that can be attained by the algorithms.

Traditionally the square norm $\nrm{\cdot}$ is used and
the stopping criterion thus requires the sum
\equation{
  \nrm{\varrho}^2=\sum_x\nrm{\varrho(x)}^2,\qquad
  \varrho(x)=\eta(x)-D\tilde{\psi}(x),
  \enum
}
to be small. Since the contributions $\nrm{\eta(x)}$ to the
norm of the source fields in lattice QCD simulations are
all about equally distributed, the square norm
$\nrm{\eta}^2$ is typically a huge
number on the order of the number $V$ of lattice points. The bound
(5.2) then does not exclude large local imbalances of the
accuracy of the approximate solution $\tilde{\psi}(x)$, where,
in the worst case, $\nrm{\varrho(x)}^2=\rmO(V)$ at some points $x$.

Inaccuracies of the solutions of the Dirac equation propagate
to the force terms in the molecular-dynamics equations
and thus to the evolved momentum and gauge fields.
Whether large local inaccuracies do occur with some appreciable
probability is currently not known, nor are their effects
on the fields and the energy deficits $\Delta H$.
Their possible presence however implies a loss of control over
the correctness of the simulations, particularly so if very large
lattices are simulated.

Unbalanced inaccuracies are excluded if the uniform norm
\equation{
  \inrm{\psi}=\sup_x\nrm{\psi(x)}
  \enum
}
is used in the stopping criterion (5.2). The uniform norm
has all properties a norm in a complex linear space must have.
Moreover,
ample empirical evidence suggests that the iterative algorithms
commonly used in lattice QCD are able to deliver accurate solutions
of the Dirac equation satisfying the uniform-norm stopping criterion
(see appendix B for further details about the uniform norm
and the associated operator norm).

\subsection 5.3 Precision loss at small quark masses

The deviation of the exact solution $\psi$ of the Dirac equation (5.1)
from any approximate solution $\tilde{\psi}$ satisfying the stopping
criterion (5.2) is bounded by [\ref{Saad}]
\equation{
  \|\tilde{\psi}-\psi\|\leq\omega\kappa(D)\|\psi\|,
  \qquad
  \kappa(D)=\|D\|\|D^{-1}\|.
  \enum
}
At small quark masses $m$
and lattice spacings $a$, the condition number $\kappa(D)$ of
the Dirac operator diverges approximately like $(am)^{-1}$ and
reaches values of $10^3$ or even $10^4$ in practice.
Since the computation of the forces in the molecular-dynamics step
requires the normal Dirac
equation, $\DD\psi=\eta$, to be solved,
the errors propagated to the momentum and the gauge field
tend to be enhanced by a factor proportional to $(am)^{-2}$.
At the end of the molecular-dynamics evolution,
the energy deficit $\Delta H$ is then obtained with an
absolute numerical error growing like $V^{1/2}(am)^{-2}$
at small $am$ and large $V$.

A reduction of these potentially catastrophic errors is automatically achieved
if the Dirac operator is split in several factors
and a corresponding number of pseudo-fermion fields is used
to represent the determinants of the factors
[\ref{Hasenbusch}--\ref{RHMCII}].
The effect is particularly transparent in the case of
the twisted-mass factorization [\ref{Hasenbusch},\ref{HasenbuschJansen}]
\equation{
  \DD=(\DD+\mu_n^2)
  \prod_{k=0}^{n-1}{\DD+\mu_k^2\over
  \DD+\mu_{k+1}^2},
  \qquad \mu_n>\mu_{n-1}>\ldots>\mu_0=0.
  \enum
}
If the largest mass, $\mu_n$, is set to a number of order $1$ in lattice
units, the computation of the force deriving from the first factor
is numerically unproblematic.
The actions associated with the rational factors, on the other hand,
may be written in the form
\equation{
  S_{{\rm pf},k}(U,\phi_k)=(\phi_k,\phi_k)+
  (\mu_{k+1}^2-\mu_k^2)(\phi_k,(D^{\dagger}\kern-1pt D+\mu_k^2)^{-1}\phi_k),
  \enum
}
which shows that the part involving the inverse of the Dirac
operator is suppressed by an explicit twisted-mass factor.
In particular, if a log-scale factorization is chosen
[\ref{TMopenQCD}], where $\mu_1$ is on the order of the quark mass
and $\mu_{k+1}\simeq 10\times\mu_k$, the twisted-mass factor
provides a suppression equal to $\mu_{k+1}^2$ of
the force deriving from eq.~(5.7) and its absolute numerical error.

\subsection 5.4 Synthesis

If all lattice sums are performed with quadruple precision,
the approximate solution of the Dirac equation is the dominant
source of numerical errors in the simulations.
The calculated values of the energy deficit
$\Delta H$ are directly affected
by these inaccuracies, but also indirectly through the
inaccuracies of the fields accumulated in
the field-rotation and molecular-dynamics steps.

Provided uniform-norm stopping criteria are used in these steps,
and if a log-scale factorization of the light-quark determinant
is chosen, the total error of $\Delta H$ is expected to
be reliably controlled by the solver tolerances.
The associated sensitivities and tolerances
can then be determined empirically so as to meet the precision
requirement discussed in sect.~4. Since the error of $\Delta H$ grows
with the number $V$ of lattice points
(roughly like $V^{1/2}$),
the tolerances must be tightened with increasing lattice size,
but there is still ample room for lattices larger than the ones
simulated to date before the tolerances reach the limit
set by the machine precision.

\section 6. Numerical studies

The simulations reported in this section mainly serve to
check whether the modified $\rmO(a)$-improved
Wilson--Dirac operator introduced
in sect.~2 is an attractive choice of
Dirac operator for numerical lattice QCD.
This is also the first time
the SMD algorithm (in the form described here with
continuous updates of the pseudo-fermion fields)
is used to simulate QCD with light quarks.
For these rather technical studies, where a direct comparison
with previous work using the standard setup is
desirable, the traditional rather than the master-field
simulation strategy is chosen.

\subsection 6.1 Non-perturbative $\rmO(a)$-improvement

As already mentioned,
the theory with tree-level improved gauge action
[\ref{TreeSymanzik}] and $2+1$ flavours of quarks,
referred to as the up, down and strange quark,
is considered.
From now on the alternative form of the improved Wilson--Dirac operator will
be assumed unless stated otherwise.
The complete specification of the theory then still requires the parameter
$\csw$ in eq.~(2.5) to be determined as a function of the bare
gauge coupling $g_0$.

The results of a non-perturbative computation of
the parameter along the lines of ref.~[\ref{cswI}]
are plotted in fig.~2.
No particular difficulty was met in this
calculation, where $\csw$ is determined by imposing
$\rmO(a)$-improvement in physically small volumes.
In the range $6/g_0^2\geq3.8$ of the coupling,
the rational function
\equation{
  \csw={1-0.325022\,g_0^2-0.0167274\,g_0^4\over1-0.489157\,g_0^2}
  \enum
}
provides an excellent fit of the data.

\topinsert
\vbox{
\vskip0ex

\epsfxsize=8.0cm\hskip2.0cm\epsfbox{plots/csw.eps}

\vskip2.0ex
\figurecaption{%
Non-perturbatively determined values (black diamonds) of the
parameter $\csw$ of the modified lattice Dirac operator
[cf.~eq.~(2.5)].
The solid curve represents the rational fit (6.1) of the data.
Also shown are the values (open squares) and fit curve (dashed line)
in the standard
$\rmO(a)$-improved theory obtained in ref.~[\ref{CswThree}].
}
}
\vskip0.0ex
\endinsert

The computation of $\csw$ was pushed up to values of the coupling,
where the lattice spacing $a$ in physical units gets close to $0.1$ fm
and higher-order lattice effects become non-negligible (scale setting
is discussed below).
Apart from a shift of the coupling to smaller values for
a given lattice spacing,
the dependence of $\csw$ on the coupling
looks similar to the one previously found in the case of the standard
$\rmO(a)$-improved Wilson--Dirac operator.

\subsection 6.2 Physical observables

Following refs.~[\ref{CLSI},\ref{CLSII}],
the renormalized parameters of the lattice theory
are taken to be the reference gradient-flow time $t_0$
[\ref{WilsonFlow}],
the pion mass $\Mpi$ and the kaon mass $\MK$.
The continuum limit, for example, is approached at fixed
\equation{
  \phi_2=8t_0\Mpi^2\quad\hbox{and}\quad
  \phi_4=8t_0\left(\MK^2+\frac{1}{2}\Mpi^2\right),
  \enum
}
while $t_0/a^2$ goes to infinity.

The use of the reference
flow time is suggested for various technical reasons,
including the fact that $t_0$ is nearly independent
of the bare light-quark mass $m_{0,u}=m_{0,d}$ when
the gauge coupling and the sum $m_{0,u}+m_{0,d}+m_{0,s}$
of the quark masses are held fixed. Along these curves
in parameter space (referred to as ``chiral trajectories''),
$\phi_4$ is nearly constant too and only $\phi_2$ varies
approximately linearly with the light-quark mass.
In particular, $\phi_4\simeq1.12$ on the trajectories passing through
the physical point [\ref{CLSI},\ref{CLSII}].
For the conversion of lattice to physical units,
the physical value of the gradient-flow smoothing range
at the reference flow time,
$\sqrt{8t_0}=0.415(4)(2)\,\fm$,
determined by Bruno et al.~[\ref{CLSII}] will be used.

Further observables considered are
the pion and kaon decay
constants\kern1pt\footnote{$\dagger$}{\footnotefont%
The normalization convention for the decay constants is
the one often used in
chiral perturbation theory, i.e.~the one where the physical value
of the pion decay constant is about $93\,\MeV$.}
\equation{
  \Fpi=\ZAh^{ud}F_{0,\pi},\qquad
  \FK=\ZAh^{us}F_{0,K},
  \enum
}
where the renormalization factors $\ZAh^{ud}$ and $\ZAh^{us}$
are computed on the same lattices as the
bare decay constants, $F_{0,\pi}$ and $F_{0,K}$,
by probing the PCAC relation at positive gradient-flow time [\ref{QuarkFlow}].
Since the calculation includes
the mass corrections required for $\rmO(a)$-improvement
[\ref{SFimp},\ref{ImpNonD}], the values of the renormalization constants
are slightly dependent on the flavour channel.
The bare decay constants and the renormalization constants
also depend on the axial-current improvement coefficient $\cA$
[\ref{SFimp}], but, as explained in appendix C, the renormalized
decay constants (6.3) are insensitive to the value
of $\cA$.

\topinsert
\newdimen\digitwidth
\setbox0=\hbox{\rm 0}
\digitwidth=\wd0
\catcode`@=\active
\def@{\kern\digitwidth}
\tablecaption{Lattice and run parameters}
\vskip-3.5ex

$$\vbox{\settabs\+&%
                  xxxxx&x&%
                  xxxxxxx&x&%
                  xxxxx&x&%
                  xxxxxxxxx&&%
                  xxxxxxxxx&&%
                  xxxxxxxxx&&%
		  xxx&&%
                  xxxxxxxx&&%
                  xxxxxxx&x\cr
\thicktablerule
\vskip1.2ex
                \+& \hfill Run\hfill
                 && \hfill Lattice\hfill
                 && \hfill $\beta$\hfill
                 && \hfill $\kappa_u$\hfill
                 && \hfill $\kappa_s$\hfill
                 && \hfill $\csw$\hfill
		 && \hfill $D$\hfill
                 && \hfill $\langle\Pacc\rangle$\hfill
                 && \hfill $N_{\rm cycles}$\hfill
                 &\cr
\vskip0.8ex
\thintablerule
\vskip1.2ex
{\ninepoint
  \+& \hfill \Ai\hfill
  &&  \hfill $96\times32^3$\hfill
  &&  \hfill $3.8$\hfill
  &&  \hfill $0.1389630$\hfill
  &&  \hfill $0.1389630$\hfill
  &&  \hfill $1.955242$\hfill
  &&  \hfill m\hfill
  &&  \hfill $0.975$\hfill
  &&  \hfill $10000$\hfill
&\cr
  \+& \hfill \Aii\hfill
  &&  \hfill $96\times32^3$\hfill
  &&  \hfill $3.8$\hfill
  &&  \hfill $0.1391874$\hfill
  &&  \hfill $0.1385164$\hfill
  &&  \hfill $1.955242$\hfill
  &&  \hfill m\hfill
  &&  \hfill $0.986$\hfill
  &&  \hfill $10000$\hfill
&\cr
  \+& \hfill \Aiii\hfill
  &&  \hfill $96\times32^3$\hfill
  &&  \hfill $3.8$\hfill
  &&  \hfill $0.1392888$\hfill
  &&  \hfill $0.1383160$\hfill
  &&  \hfill $1.955242$\hfill
  &&  \hfill m\hfill
  &&  \hfill $0.982$\hfill
  &&  \hfill $10000$\hfill
&\cr
  \+& \hfill \Bi\hfill
  &&  \hfill $96\times48^3$\hfill
  &&  \hfill $4.0$\hfill
  &&  \hfill $0.1382720$\hfill
  &&  \hfill $0.1382720$\hfill
  &&  \hfill $1.783303$\hfill
  &&  \hfill m\hfill
  &&  \hfill $0.988$\hfill
  &&  \hfill $24000$\hfill
&\cr
  \+& \hfill \Ci\hfill
  &&  \hfill $96\times32^3$\hfill
  &&  \hfill $@3.36$\hfill
  &&  \hfill $0.1366400$\hfill
  &&  \hfill $0.1366400$\hfill
  &&  \hfill $2.038765$\hfill
  &&  \hfill t\hfill
  &&  \hfill $0.996$\hfill
  &&  \hfill $10000$\hfill
&\cr
}
\vskip1.0ex
\thicktablerule
}
$$
\vskip0.0ex
\endinsert

\subsection 6.3 Simulation algorithm

Apart from the use of the SMD in place of the HMC algorithm,
the simulations reported in this paper were set up as
described in ref.~[\ref{TMopenQCD}].
In particular,
twisted-mass factorizations were chosen
for the light-quark determinant,
rational approximations for the strange-quark determinant
and a hierarchical 4th-order integrator with 2 levels
for the numerical integration of the molecular-dynamics equations.
Following the recommendations in sect.~5, the uniform-norm
stopping criterion was used in the computations of the forces
deriving from the pseudo-fermion actions, with tolerances far
sufficient for the numerical errors of $\Delta H$ to be
negligible.

All simulations were
performed using the publicly available {\tt openQCD}
program package [\ref{OQCD}].
The SMD friction parameter $\gamma$ was set to $0.3$, as suggested
in ref.~[\ref{OpenBC}],
and the SMD step size $\eps=0.31$ was chosen so that
a high acceptance rate is attained with 2 steps at
the outer level of the molecular-dynamics integration scheme
(3 steps in the case of run \Ci).

\topinsert
\newdimen\digitwidth
\setbox0=\hbox{\rm 0}
\digitwidth=\wd0
\catcode`@=\active
\def@{\kern\digitwidth}
\tablecaption{Lattice spacing, lattice size,
masses and decay constants in physical units}
\vskip-3.5ex

$$\vbox{\settabs\+&%
                  xxxxx&x&%
                  xxxx&x&%
                  xxxxxxx&&%
                  xxxxxxxxxi&x&%
                  xxxxxxxxxi&x&%
                  xxxxxxxxxi&x&%
                  xxxxxxxxxi&x&%
                  xxxxxx&\cr
\thicktablerule
\vskip1.2ex
                \+& \hfill Run\hfill
                 && \hfill $a\,[\fm]$\hfill
                 && \hfill $L\,[\fm]$\hfill
                 && \hfill $\Mpi\,[\MeV]$\hfill
                 && \hfill $\MK\,[\MeV]$\hfill
                 && \hfill $\Fpi\,[\MeV]$\hfill
                 && \hfill $\FK\,[\MeV]$\hfill
                 && \hfill $\Mpi L$\hfill
                 &\cr
\vskip0.8ex
\thintablerule
\vskip1.2ex
{\ninepoint
  \+& \hfill \Ai\hfill
  &&  \hfill $0.094$\hfill
  &&  \hfill $3.0$\hfill
  &&  \hfill $408$\hfill
  &&  \hfill $-$\hfill
  &&  \hfill $108$\hfill
  &&  \hfill $-$\hfill
  &&  \hfill $6.2$\hfill
&\cr
  \+& \hfill \Aii\hfill
  &&  \hfill $0.094$\hfill
  &&  \hfill $3.0$\hfill
  &&  \hfill $293$\hfill
  &&  \hfill $454$\hfill
  &&  \hfill $100$\hfill
  &&  \hfill $109$\hfill
  &&  \hfill $4.5$\hfill
&\cr
  \+& \hfill \Aiii\hfill
  &&  \hfill $0.093$\hfill
  &&  \hfill $3.0$\hfill
  &&  \hfill $215$\hfill
  &&  \hfill $470$\hfill
  &&  \hfill $@96$\hfill
  &&  \hfill $109$\hfill
  &&  \hfill $3.3$\hfill
&\cr
  \+& \hfill \Bi\hfill
  &&  \hfill $0.064$\hfill
  &&  \hfill $3.1$\hfill
  &&  \hfill $409$\hfill
  &&  \hfill $-$\hfill
  &&  \hfill $107$\hfill
  &&  \hfill $-$\hfill
  &&  \hfill $6.4$\hfill
&\cr
  \+& \hfill \Ci\hfill
  &&  \hfill $0.095$\hfill
  &&  \hfill $3.0$\hfill
  &&  \hfill $408$\hfill
  &&  \hfill $-$\hfill
  &&  \hfill $107$\hfill
  &&  \hfill $-$\hfill
  &&  \hfill $6.3$\hfill
&\cr
}
\vskip1.0ex
\thicktablerule
}
$$
\vskip0.0ex
\endinsert

\subsection 6.4 Lattice parameters

The basic parameters of the simulated lattices are listed in table~1,
as usual quoting the values of $\beta=6/g_0^2$
and the hopping parameters $\kappa=1/(8+2m_0)$
instead of the bare coupling and masses.
Periodic boundary conditions (anti-periodic in time for the quark
fields) were imposed in all cases and, as indicated
by the entries in the column labeled ``$D$'',
run \Ci{} is the only one where the traditional form of the Dirac operator
was used.
In the last two columns,
the acceptance rates and numbers of SMD update cycles performed
after thermalization are listed.

\topinsert
\newdimen\digitwidth
\setbox0=\hbox{\rm 0}
\digitwidth=\wd0
\catcode`@=\active
\def@{\kern\digitwidth}
\tablecaption{Masses and decay constants in units of $t_0$}
\vskip-3.5ex

$$\vbox{\settabs\+&%
                  xxxx&&%
                  xxxxxxxxxxx&&%
                  xxxxxxxxxxx&&%
                  xxxxxxxxxxx&x&%
                  xxxxxxxxxxx&x&%
                  xxxxxxxxxxx&\cr
\thicktablerule
\vskip1.2ex
                \+& \hfill Run\hfill
                 && \hfill $t_0/a^2$\hfill
                 && \hfill $\phi_2$\hfill
                 && \hfill $\phi_4$\hfill
                 && \hfill $\sqrt{8t_0}\Fpi$\hfill
                 && \hfill $\sqrt{8t_0}\FK$\hfill
                 &\cr
\vskip0.8ex
\thintablerule
\vskip1.2ex
{\ninepoint
  \+& \hfill \Ai\hfill
  &&  \hfill $2.4420(36)$\hfill
  &&  \hfill $0.7366(39)$\hfill
  &&  \hfill $1.1049(45)$\hfill
  &&  \hfill $0.22701(60)$\hfill
  &&  \hfill $-$\hfill
&\cr
  \+& \hfill \Aii\hfill
  &&  \hfill $2.4540(27)$\hfill
  &&  \hfill $0.3807(40)$\hfill
  &&  \hfill $1.1056(51)$\hfill
  &&  \hfill $0.21095(72)$\hfill
  &&  \hfill $0.23018(50)$\hfill
  &\cr
  \+& \hfill \Aiii\hfill
  &&  \hfill $2.4645(35)$\hfill
  &&  \hfill $0.2039(56)$\hfill
  &&  \hfill $1.0873(60)$\hfill
  &&  \hfill $0.20119(86)$\hfill
  &&  \hfill $0.22912(56)$\hfill
  &\cr
  \+& \hfill \Bi\hfill
  &&  \hfill $5.2470(57)$\hfill
  &&  \hfill $0.7414(39)$\hfill
  &&  \hfill $1.1121(44)$\hfill
  &&  \hfill $0.22430(66)$\hfill
  &&  \hfill $-$\hfill
  &\cr
  \+& \hfill \Ci\hfill
  &&  \hfill $2.3952(34)$\hfill
  &&  \hfill $0.7361(76)$\hfill
  &&  \hfill $1.1042(85)$\hfill
  &&  \hfill $0.22452(55)$\hfill
  &&  \hfill $-$\hfill
  &\cr
}
\vskip1.0ex
\thicktablerule
}
$$
\vskip0.0ex
\endinsert

The simulated lattices are in the large-volume regime of QCD
and approximately on the physical chiral trajectories $\phi_4\simeq1.12$
[\ref{CLSI},\ref{CLSII}]
(see tables~2 and 3). In the case of the lattices \Ai--\Aiii,
only the bare quark masses change, keeping their sum
exactly constant, while \Ai, \Bi{} and \Ci{}
are all at the $\SUthree$-symmetric point. In particular,
the kinematical situation on the \Ai{} and \Ci{} lattices is
practically the same.
\Ai{} and \Bi, on the other hand, allow the
behaviour of the simulations and the lattice effects
to be studied as the lattice spacing is reduced at fixed meson masses
(i.e.~fixed $\phi_2$ and $\phi_4$).
The figures listed in table~2 serve for illustration
only and are therefore given without errors.

\subsection 6.5 Simulation stability

As previously noted, large energy violations $\Delta H$
in the molecular-dynamics step of the SMD algorithm
are a sign of potentially harmful algorithmic instabilities.
On the coarser lattices simulated, \Ai--\Aiii{} and \Ci,
the fractions $\rspike$ of update cycles where $|\Delta H|>1$
(i.e.~where $\Delta H$ has a ``spike'') are one or
two per mille and thus fairly small (see table~4).
When the lattice spacing is reduced, the stability of the
molecular-dynamics integration improves and not a single
spike was seen in run \Bi.

While the frequency of spikes is similar in run \Ai{} and \Ci,
the molecular-dynamics equations had to be integrated
with a $1.5$ times smaller step size
in run \Ci{} to achieve this.
The use of the traditional instead of the alternative
form of the Dirac operator also leads to larger fluctuations and
a much longer integrated autocorrelation time $\tau(\pbar)$ of the
average plaquette $\pbar$ (see fig.~3 and table~4).
Since the distribution of the latter is a property of the theory,
these large fluctuations cannot be attributed to
the choice of simulation algorithm and
may instead signal the onset of a change of regime
at this coarse lattice spacing.
The observed autocorrelations of the average plaquette
however do suggest the same, given that this quantity is dominated by
the high-frequency modes of the gauge field,
which are normally rapidly equilibrated by the SMD algorithm.
On finer lattices the effect in fact
quickly disappears.

\subsection 6.6 Spectral gap

The correctness and efficiency of the simulations
may also be negatively affected if the
lattice Dirac operator does not have a safe spectral gap
[\ref{Stability}].
Compared to the ones previously
obtained in two-flavour QCD [\ref{Stability},\ref{StabilityII}],
the distributions of the lowest eigenvalue
of the light-quark operator $(\DD)^{1/2}$ measured in the runs
\Ai--\Ci{} turn out to be rather similar (see table~4 and appendix D).
In particular, the distributions are separated from the origin
on all these lattices.
Their medians $\bar{\mu}$ (fourth column in table~4)
are lower than the product of the axial-current
renormalization constant times the bare current-quark mass
$\frac{1}{2}m_{ud}$ of the light quarks (last column),
but the difference decreases toward the continuum limit
as has to be the case [\ref{Stability}].

\topinsert
\newdimen\digitwidth
\setbox0=\hbox{\rm 0}
\digitwidth=\wd0
\catcode`@=\active
\def@{\kern\digitwidth}
\tablecaption{Observables related to the simulation stability}
\vskip-3.5ex

$$\vbox{\settabs\+&%
                  xxxx&x&%
                  xxxxxxxx&x&%
                  xxxxxxxxx&x&%
                  xxxxxxxxx&x&%
                  xxxxxxxxx&x&%
                  xxxxxxxxx&x&%
                  xxxxxxxxxxxxxx&\cr
\thicktablerule
\vskip1.2ex
                \+& \hfill Run\hfill
                 && \hfill $\rspike\,[\%]$\hfill
                 && \hfill $\tau(\pbar)$\hfill
                 && \hfill $\evm\times10^3$\hfill
                 && \hfill $\evs\times10^3$\hfill
                 && \hfill $\evs\sqrt{V}$\hfill
                 && \hfill $\frac{1}{2}\ZAh^{ud}m_{ud}\times10^3$\hfill
                 &\cr
\vskip0.8ex
\thintablerule
\vskip1.2ex
{\ninepoint
  \+& \hfill \Ai\hfill
  &&  \hfill $0.19(10)$\hfill
  &&  \hfill $7.6(1.6)$\hfill
  &&  \hfill $5.251(89)$\hfill
  &&  \hfill $0.613(57)$\hfill
  &&  \hfill $1.09(10)@$\hfill
  &&  \hfill $8.508(34)$\hfill
  &\cr
  \+& \hfill \Aii\hfill
  &&  \hfill $0.19(10)$\hfill
  &&  \hfill $5.4(1.1)$\hfill
  &&  \hfill $2.283(51)$\hfill
  &&  \hfill $0.399(32)$\hfill
  &&  \hfill $0.708(57)$\hfill
  &&  \hfill $4.260(32)$\hfill
  &\cr
  \+& \hfill \Aiii\hfill
  &&  \hfill $0.10(7)@$\hfill
  &&  \hfill $7.4(1.4)$\hfill
  &&  \hfill $1.218(46)$\hfill
  &&  \hfill $0.351(33)$\hfill
  &&  \hfill $0.623(59)$\hfill
  &&  \hfill $2.260(50)$\hfill
  &\cr
  \+& \hfill \Bi\hfill
  &&  \hfill $0.0{\phantom{0(10)}}$\hfill
  &&  \hfill $3.5(4){\phantom{6.}}$\hfill
  &&  \hfill $4.541(13)$\hfill
  &&  \hfill $0.310(8)@$\hfill
  &&  \hfill $1.011(27)$\hfill
  &&  \hfill $5.901(18)$\hfill
  &\cr
  \+& \hfill \Ci\hfill
  &&  \hfill $0.22(5)@$\hfill
  &&  \hfill $>30@$\hfill
  &&  \hfill $4.071(49)$\hfill
  &&  \hfill $0.843(30)$\hfill
  &&  \hfill $1.495(53)$\hfill
  &&  \hfill $7.760(60)$\hfill
  &\cr
}
\vskip1.0ex
\thicktablerule
\vskip1.5ex
\noindent
{\footnotefont%
$\tau(\pbar)$ is given in units of molecular-dynamics time
and the values in columns $4-7$ in lattice units.}
}
$$
\vskip0.0ex
\endinsert

\topinsert
\vbox{
\vskip0ex

\epsfxsize=12.5cm\hskip-0.3cm\epsfbox{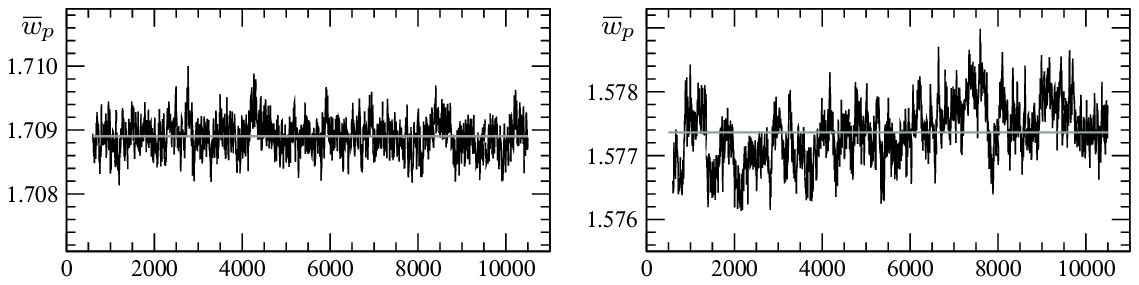}

\vskip2.0ex
\figurecaption{%
Histories of the average Wilson plaquette loop $\pbar$
versus the SMD update cycle number.
The values plotted are the ones measured
after every 5th cycle in run \Ai{} (left) and \Ci{} (right).
In both plots the same scale is used on the ordinate.
}
}
\vskip0.0ex
\endinsert

Theoretical arguments and numerical evidence produced in
refs.~[\ref{Stability},\ref{StabilityII}]
suggest that the widths $\evs$ of the eigenvalue distributions
are, in lattice units,
roughly equal to $1/\sqrt{V}$ (where $V$ denotes
the number of lattice points). The values of
$\evs\sqrt{V}$ listed in table~4 are broadly consistent with this rule
and moreover show that the product tends to
decrease with $\Mpi L$, an observation previously
made in the standard $\rmO(a)$-improved two-flavour theory
[\ref{StabilityII}]. With respect to the distribution measured in
run \Ai, the one obtained in run \Ci{} is however
noticeably wider, i.e.~at this fairly coarse lattice spacing, the
use of the modified Dirac operator leads to a narrower eigenvalue
distribution.

\subsection 6.7 Higher-order lattice effects

All simulated lattices are in a range of parameters, where
$\rmO(a^2)$ lattice effects cannot be expected to be
very small, independently of which Dirac operator is chosen. The
values of $t_0/a^2$, for example, would change by $15\%$
on the \Ai{} lattice and by $6\%$ on the \Bi{} lattice,
if defined with the Wilson plaquette
instead of the symmetric (``clover'') expression
for the Yang--Mills action density.

\topinsert
\vbox{
\vskip0ex

\epsfxsize=9.0cm\hskip1.5cm\epsfbox{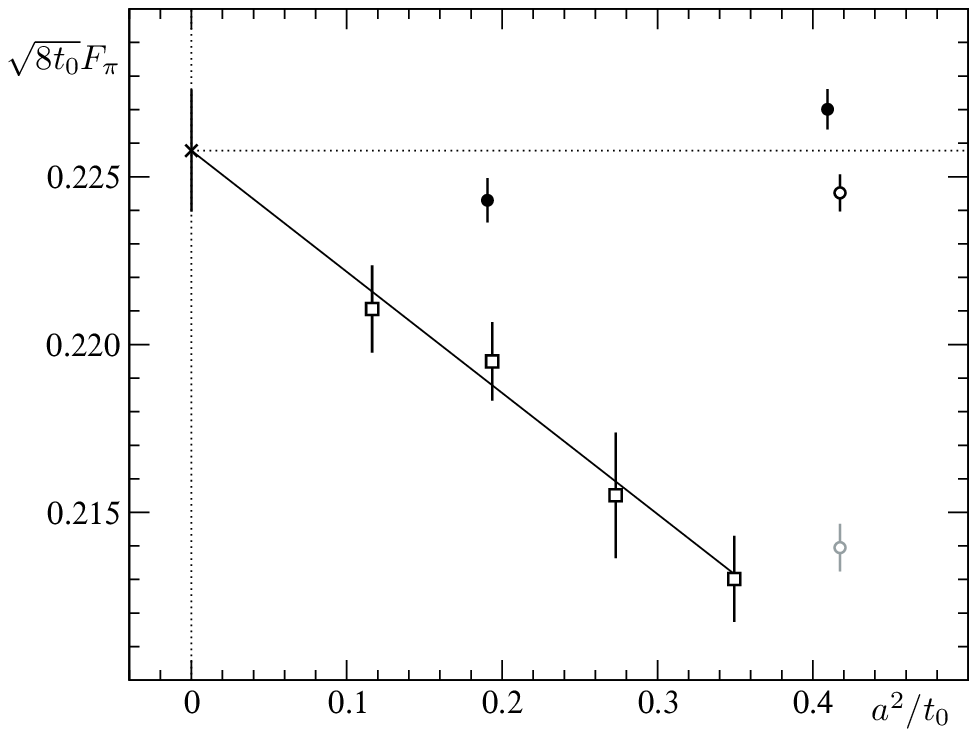}

\vskip3.0ex
\figurecaption{%
Lattice-spacing dependence of $\sqrt{8t_0}\Fpi$ at
$m_{0,u}=m_{0,d}=m_{0,s}$.
Open squares represent
results previously obtained
in ref.~[\ref{CLSII}] at $\phi_4=1.11$ using the traditional
setup of the $\rmO(a)$-improved theory.
A linear extrapolation of these data
yields a value in the continuum limit (cross), which coincides
with the results obtained in the runs \Ai{}, \Bi{} (black circles) and \Ci{}
(open circle). The latter moves down
if the axial-current renormalization constant is replaced by the one
used in ref.~[\ref{CLSII}] (grey open circle).
}
}
\endinsert

The dimensionless combination $\sqrt{8t_0}\Fpi$
is potentially more sensitive to the choice of the lattice
Dirac operator than gluonic quantities like $t_0$.
In fig.~4 the results for $\sqrt{8t_0}\Fpi$ obtained
at the $\SUthree$-symmetric point (i.e.~in the runs \Ai, \Bi{} and \Ci)
are compared with data published in ref.~[\ref{CLSII}].
For a sensible comparison, $\phi_4$ should assume
the same value on all lattices. This is not exactly the case,
but the variations in $\phi_4$ are too small to have a noticeable
effect. Different (non-perturbative) strategies to determine
the axial-current renormalization constant however lead to values
of $\Fpi$ differing by $6\%$ and more on the coarser
lattices simulated
(see fig.~4)\kern1pt\footnote{$\dagger$}{\footnotefont%
In ref.~[\ref{CLSII}] the values of $\ZA$ extracted from the so-called
chirally rotated Schr\"odinger functional [\ref{XSF}] were used.
Even smaller values
of $\ZA$ were previously found by probing a chiral Ward identity on lattices
with ordinary Schr\"odinger-functional boundary conditions [\ref{ZAThree}].}.
The chosen renormalization condition for the axial current
thus matters, and it is partly a consequence of the choice made here
that the computed values of $\sqrt{8t_0}\Fpi$ coincide with
the continuum value determined in ref.~[\ref{CLSII}] within
a margin of $0.7\%$.

\topinsert
\vbox{
\vskip0ex

\epsfxsize=12.5cm\hskip-0.30cm\epsfbox{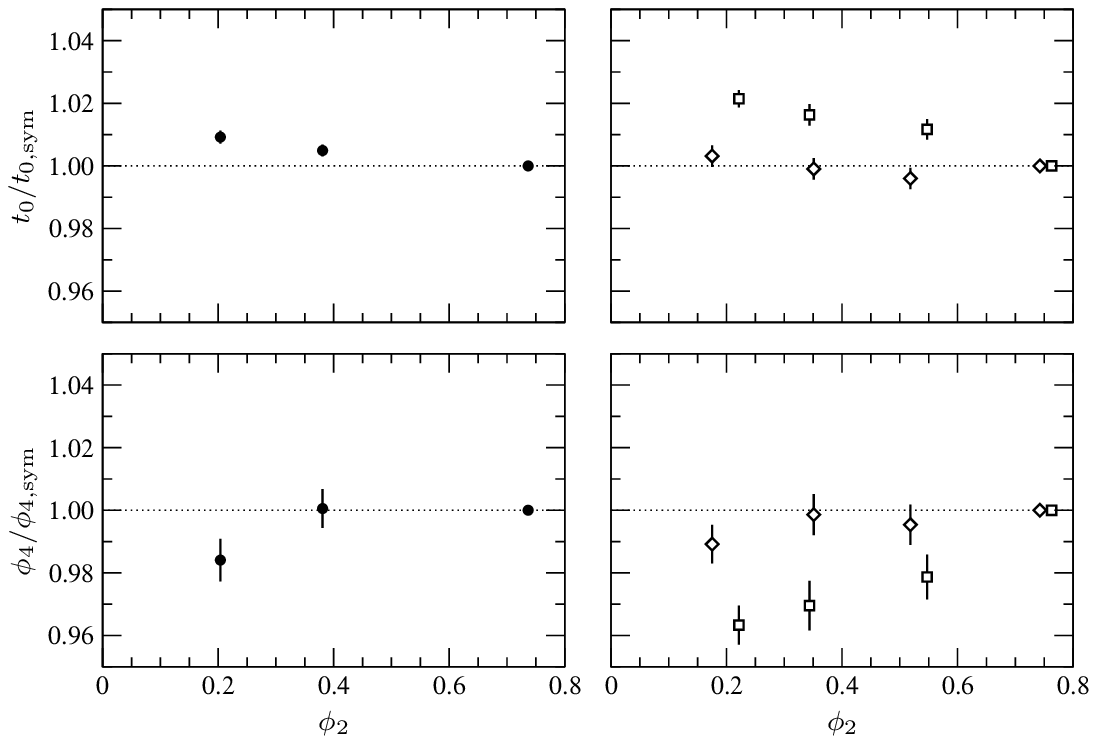}

\vskip1.0ex
\figurecaption{%
Quark-mass dependence of $t_0$ and $\phi_4$
along chiral trajectories
(approximate\-ly) passing through the physical point.
All data are normalized by the values measured
at the $\SUthree$-symmetric points on the trajectories.
The plots on the right show results reported in ref.~[\ref{CLSII}]
from simulations of two series of lattices
with spacing $a=0.086\,\fm$ (open squares) and $a=0.064\,\fm$
(open diamonds), respectively, in which the traditional
form of the improved lattice Dirac operator was used.
On the left the results obtained
in the runs \Ai--\Aiii{}, where the lattice spacing is
about $0.094\,\fm$, are displayed (black circles).
}
}
\endinsert

The quark-mass dependence of $t_0$ and $\phi_4$ along
the chiral trajectories in parameter space
provides another opportunity to study the magnitude of the
lattice effects.
As can be seen from the data plotted on the right of fig.~5, the
dependence on the mass of the light quarks becomes flatter when the
continuum limit is approached on the physical trajectories.
The corresponding
results obtained in runs \Ai--\Aiii{} (plots on the left)
thus show that the use of the modified lattice Dirac operator leads,
in the case of these observables,
to significantly reduced lattice effects.

\subsection 6.8 Miscellaneous remarks

{\it SMD cycle timing}.
Per unit of molecular-dynamics time, the SMD algorithm updates
the pseudo-fermion fields more frequently than the HMC algorithm
and the number of accept-reject steps is larger as well.
In practice this overhead however
accounts for only a small fraction of the required computer time
(about $5\%$ in the runs \Ai--\Aiii{} and \Bi{}),
if the solutions of the Dirac equation are reused
whenever this is profitable.

Algorithmic instabilities tend to have a cost in terms of computer
time too. In run \Ci, for example, the molecular-dynamics integration
step size had to be reduced with respect to the
one in run \Ai{} in order to achieve a similarly
low rate of spikes in $\Delta H$. The larger fluctuations
of the gauge field moreover made it more difficult to
obtain the solutions of the Dirac equation to the required precision.
As a result, run \Ci{} consumed about two times
the computer time spent for run \Ai.

\vskip1ex\noindent
{\it Autocorrelations}.
The simulations \Ai--\Ci{} are not long enough for
an accurate determination of the integrated autocorrelation times
of the calculated physical quantities.
Significant autocorrelations were however not
observed in the runs \Ai--\Aiii{} and \Ci{}
among measurements separated by 100 SMD update cycles,
while practically decorrelated measurements were obtained
in run \Bi{} at two times larger separations.
The behaviour is thus largely the same
as in the more extensive SMD simulations of the $\SUthree$ gauge theory
reported in ref.~[\ref{OpenBC}].

\vskip1ex\noindent
{\it Statistical errors of $\Fpi$ and $\FK$}.
Although the available statistics is rather limited,
the renormalized pseudo-scalar decay constants are obtained with
statistical errors of a fraction of a percent (cf.~table~3).
The smallness of these errors partly derives from the fact
that the axial-current renormalization constants are
determined on the same lattices as the bare decay constants.
At the smaller light-quark masses, in particular,
the error correlations are then such that the renormalized
decay constants are obtained with significantly better
statistical precision than the unrenormalized ones.

\vskip1ex\noindent
{\it Simulations with open boundary conditions}.
As a further check on the viability of the modified
lattice Dirac operator,
a $96\times48^3$ lattice with open boundary conditions in time
[\ref{OpenBC}] was simulated at $\beta=4.1$.
The run turned out to be completely stable, with zero spikes in $\Delta H$,
and no indication of an unexpected behaviour triggered by the presence
of the boundaries of the lattice.

\section 7. Conclusions

The alternative $\rmO(a)$-improved Wilson--Dirac operator
and the algorithmic measures put forward in this paper
have a stabilizing effect on lattice QCD simulations, particularly
so on coarse lattices. With respect to the traditional setup,
there do not appear to be any disadvantages.
Non-perturbative improvement, for example, works out in much the
same way and the observed residual lattice effects
turned out to be smaller.
Moreover, the simulations tend to be somewhat faster.

Another interesting outcome, unrelated to the stability issues,
concerns the renormalization of the axial currents.
On coarse lattices, different renormalization conditions
can lead to significantly different results for the renormalized
pseudo-scalar decay constants.
Here the current renormalization factors were determined by probing
chiral symmetry at positive gradient-flow time [\ref{QuarkFlow}]
and, with this choice, the calculated decay constants
turned out to depend only weakly on the lattice spacing
(see fig.~4).
The renormalization procedure has further technical advantages,
among them the fact that
the computation of the renormalization factors does not require
additional simulations and that
the renormalized decay constants can be shown to be insensitive to the value
of the axial-current improvement coefficient $\cA$.

While master-field simulations with light quarks await to be performed,
there is every reason to expect such simulations to work out if the
recommendations given in this paper are followed. Making most out
of such simulations is however a non-trivial task.
The evaluation of hadron propagators, for example,
and their statistical errors will have
to be reconsidered [\ref{LuscherMaster}].

\vskip1ex
{\ninepoint
P.F.~thanks the Institute of Theoretical Physics at the WWU M\"unster
for hospitality and Jochen Heitger for useful discussions.
Thanks also go to the ALPHA collaboration for supporting this work
and to Mattia Bruno for providing a table of
the CLS data plotted in fig.~4.
A.R.~is supported in part by the STFC Consolidated Grant ST/P000479/1.
Many simulations reported in this paper were performed on a dedicated
HPC cluster at CERN.
This work also used the DiRAC Extreme Scaling service at the University of
Edinburgh, operated by the Edinburgh Parallel Computing Centre on behalf of
the STFC DiRAC HPC Facility (www.dirac.ac.uk) and funded by
BEIS capital funding via STFC capital grant ST/R00238X/1 and STFC DiRAC
Operations grant ST/R001006/1. DiRAC is part of the UK National
e-Infrastructure. The generous support of all these institutions is gratefully
acknowledged.
}

\appendix A. Implementation of the modified improved Dirac operator

If a chiral representation of
the Dirac matrices is chosen, the
Pauli term assumes the block-diagonal form
\equation{
  {\csw\over4+m_0}
  \sum_{\mu,\nu=0}^3\frac{i}{4}
  \sigma_{\mu\nu}\widehat{F}_{\mu\nu}=
  \pmatrix{A_{+} & 0\cr
           \noalign{\vskip1ex}
           0 & A_{-}\cr}
  \enum
}
and the diagonal part (2.5) of the modified Dirac operator
is then given by
\equation{
  \Dee+\Doo=\hbox{constant}\times\pmatrix{\rme^{A_{+}} & 0\cr
                                         \noalign{\vskip1ex}
                                         0 & \rme^{A_{-}} \cr}.
  \enum
}
The blocks $A_{\pm}$ in these equations are
traceless Hermitian $6\times6$ matrices
acting on the upper and lower two Dirac components of the quark fields,
respectively.

Since the Pauli term already appears in the traditional form of
the O($a$)-improved Dirac operator, its numerical evaluation and
differentiation with respect to the gauge field are not discussed here.
Instead some efficient methods to compute the exponential $\rme^A$ and
its differential $\rmd\rme^A$ with respect to the parameters of
$A$ are described, where $A$ stands for either $A_{+}$ or $A_{-}$.

\subsection A.1 Evaluation of the exponential function

The Cayley--Hamilton theorem
asserts that any matrix satisfies its own
characteristic equation. Specifically, in the case considered here,
the equation
\equation{
  A^6=\sum_{k=0}^4p_kA^k
  \enum
}
holds, with real coefficients $p_k$ given by
\equation{
  p_0=\frac{1}{6}\tr\{A^6\}
      -\frac{1}{8}\tr\{A^4\}\tr\{A^2\}-\frac{1}{18}\tr\{A^3\}^2
      +\frac{1}{48}\tr\{A^2\}^3,
  \enum
  \next{2.0ex}
  p_1=\frac{1}{5}\tr\{A^5\}-\frac{1}{6}\tr\{A^3\}\tr\{A^2\},
  \enum
  \next{2.0ex}
  p_2=\frac{1}{4}\tr\{A^4\}-\frac{1}{8}\tr\{A^2\}^2,
  \enum
  \next{2.0ex}
  p_3=\frac{1}{3}\tr\{A^3\},
  \enum
  \next{2.0ex}
  p_4=\frac{1}{2}\tr\{A^2\}.
  \enum
}
An arbitrary polynomial $\sum_{k=0}^Nc_kA^k$
in $A$ of degree $N\geq6$ can therefore be reduced to a polynomial
of degree $5$ with $A$-dependent coefficients $d_0,\ldots,d_5$, which may
be calculated through a simple recursion.

The bound (2.4) implies that the eigenvalues $\lambda$ of $A$
satisfy
\equation{
  |\lambda|\leq R,\qquad R={3\csw\over4+m_0}.
  \enum
}
At the values of $\csw$ and $m_0$ of interest,
the Taylor series
\equation{
  \rme^A=\sum_{k=0}^N{A^k\over k!}+r_N(A)
  \enum
}
is thus guaranteed to converge rapidly, with a remainder bounded by
\equation{
  \|r_N(A)\|_2\leq{R^{N+1}\over(N+1)!}\rme^R.
  \enum
}
Using the Cayley--Hamilton
theorem, the exponential of the Pauli
term can in this way easily be obtained to machine precision.

\subsection A.2 Differential of the exponential function

The differential of $\rme^A$
with respect to the independent parameters of $A$ is given by
\equation{
  \rmd\rme^A=\int_0^1\rmd t\,\rme^{tA}\rmd A\,
  \rme^{(1-t)A}.
  \enum
}
Expansion of the exponentials on the right then yields the
rapidly convergent series
\equation{
  \rmd\rme^A=\sum_{k=0}^N\sum_{l=0}^{N-k}
  {1\over(k+l+1)!}A^k\rmd A A^l+r_N(A,\rmd A),
  \enum
}
where
\equation{
  \|r_N(A,\rmd A)\|_2\leq{R^{N+1}\over(N+1)!}
  \rme^R\|\rmd A\|_2.
  \enum
}
The derivatives of the exponential can thus be obtained to machine precision
by truncating the series at the same value of $N$ as the series (A.10).
Application of the Cayley--Hamilton theorem and the recursion
that derives from it finally leads to the expression
\equation{
  \rmd\rme^A=\sum_{k,l=0}^5 C_{kl}A^k\rmd A A^l+r_N(A,\rmd A),
  \enum
}
$C$ being a real symmetric $6\times6$ matrix
that depends on $N$ and $A$.

In practice only matrix elements $(v,\rmd\rme^A w)$ of the differential
need to be computed, where $v$ and $w$ are the upper
or lower two Dirac components of quark spinors at
a given lattice point. The evaluation
of these matrix elements proceeds by computing the spinors
$A^kv$ and $A^lw$ for all $k,l=0,\ldots,5$, then the linear
combinations $\sum_lC_{kl}A^lw$ and finally the scalar
products of these with $A^kv$.

\appendix B. Properties of the uniform norm

In this appendix, some basic facts about
the uniform norm (5.4) are briefly described,
omitting the elementary but often lengthy proofs of
the statements made.

\subsection B.1 Uniform norm of random fields

Let $\eta$ be a random quark field with normal distribution.
Its uniform norm, $r=\inrm{\eta}$, is a random variable,
whose distribution $p(r)$ on a lattice with $V$ points,
\equation{
  p(r)={\rmd\over\rmd r}\left\{f(r)^V\right\},
  \qquad
  f(r)={2\over 11!}\int_0^r\rmd s\,s^{23}\rme^{-s^2},
  \enum
}
has an approximately Gaussian shape.
When $V$ increases from, say, $10^4$ to $10^{12}$, the position
of the maximum of the distribution slowly moves from
$5.42$ to $7.37$. Since the probability for $\inrm{\eta}$
to be larger than these values rapidly gets extremely small,
the uniform norm of random fields is, in practice, typically
in the range from $5$ to $8$.

\subsection B.2 Associated operator norm

The uniform norm of a linear operator $A$ acting on quark fields
$\psi(x)$ is defined by
\equation{
  \inrm{A}=\sup_{\psi\neq0}{\inrm{A\psi}
  \over\inrm{\psi}}.
  \enum
}
It has all the usual properties of an operator norm and in particular
satisfies
\equation{
  \inrm{AB}\leq\inrm{A}\inrm{B}
  \enum
}
for any pair $A,B$ of linear operators. The norm of $A$ and its Hermitian
conjugate $A^{\dagger}$ however need not be the same, a notable
exception being the $\dirac{5}$-Hermitian Wilson--Dirac operator.

In terms of the position-space kernel,
\equation{
  (A\psi)(x)=\sum_yA(x,y)\psi(y),
  \enum
}
the uniform norm is given by
\equation{
  \inrm{A}=
  \sup_x\biggl\{\sup_{\nrm{s}=1}\sum_y\nrm{A(x,y)^{\dagger}s}\biggr\},
  \enum
}
where the inner supremum is taken over all $y$-independent
spinors $s$ of norm $1$.

\subsection B.3 Uniform-norm condition number of the Dirac operator $D$

The uniform-norm condition number
\equation{
  \kappa_{\infty}(D)=\inrm{D}\inrm{D^{-1}}
  \enum
}
of the Dirac operator can be shown to be at least as large as
the condition number $\kappa_2(D)$
defined through the square norm. While the bound
\equation{
  \inrm{D}\leq\nrm{\Dee+\Doo}+4\sqrt{2}
  \enum
}
is easily derived, an estimation of the second factor in eq.~(B.6),
\equation{
  \inrm{D^{-1}}=
  \sup_x\biggl\{\sup_{\nrm{s}=1}\sum_y\nrm{S(x,y)^{\dagger}s}\biggr\},
  \enum
}
requires the long-distance behaviour of the quark propagator $S(x,y)$
to be known.
In the case of free quarks on an infinite lattice, for example,
the leading asymptotic behaviour near the chiral limit,
\equation{
  \inrm{D^{-1}}
  \mathrel{\mathop=_{m_0\to0}}
  {c_1\over m_0}+\ldots,\qquad c_1=2.60(1),
  \enum
}
turns out to be practically the same as the one of the square norm.

The spontaneous breaking of chiral symmetry
in QCD however leads to a more singular chiral behaviour.
Assuming mass-degenerate up and down quarks,
the norm of the light-quark propagator can be estimated
by replacing $\nrm{S(x,y)s}^2$ through its average value,
the pion propagator, and by using chiral perturbation theory
for the latter. The asymptotic formula
\equation{
  \inrm{D^{-1}}
  \mathrel{\mathop=_{m_{ud}\to0}}
  {c_2\Fpi\over\Mpi\ZA^{ud}m_{ud}}+\ldots,
  \qquad c_2=36.1(1),
  \enum
}
obtained in this way then shows that the condition number
$\kappa_{\infty}(D)$ must be expected to grow more rapidly than $\kappa_2(D)$
in the chiral limit, although in practice the factor $c_2\Fpi/\Mpi$
never becomes very large.

\appendix C. Axial current renormalization

The goal in this appendix is to show that the renormalized
decay constants $\Fpi$ and $\FK$ are insensitive to the value
of the axial-current improvement coefficient $\cA$ if
the bare decay constants are extracted from
the vacuum-to-meson matrix elements of the axial currents and
if the currents are renormalized as described
in ref.~[\ref{QuarkFlow}].

\subsection C.1 Masses and decay constants

In order to simplify the notation,
an unspecified non-singlet flavour channel is considered, the
flavour indices are omitted and the lattice spacing is set to
unity. Moreover, the time extent $T$ of the lattice is assumed
to be sufficiently large that its effects are completely negligible.

The bare $\rmO(a)$-improved axial current
\equation{
  (\imp{A})_{\mu}(x)=A_{\mu}(x)+\cA\ring{\partial}_{\mu}P(x)
  \enum
}
is a linear combination of the bare current $A_{\mu}(x)$
and the gradient of the axial density $P(x)$, where
$\ring{\partial}_{\mu}$ denotes the symmetric
nearest-neighbour difference operator
[\ref{SFimp}].
At large times $x_0$ and up to exponentially suppressed terms,
the two-point functions
\equation{
  \sum_{\xvec}\langle P(x)P(0)\rangle=-{G^2\over M}\,\rme^{-Mx_0}
  +\ldots,
  \enum
  \nexteq{2.5ex}
  \sum_{\xvec}\langle(\imp{A})_0(x)P(0)\rangle=
  \imp{F}G\,\rme^{-Mx_0}
  +\ldots,
  \enum
}
are given by the pseudo-scalar meson mass $M$, the vacuum-to-meson
matrix element $G$ of the axial density
and the bare improved decay constant $\imp{F}$.

Following common practice,
the sum $m$ of the current-quark masses in the chosen flavour
channel may be defined by requiring the PCAC relation
\equation{
  \sum_{\xvec}\langle
  \{\ring{\partial}_0A_0(x)+\cA\drvstar{0}\drv{0}P(x)\}P(0)\rangle=
  m\sum_{\xvec}\langle P(x)P(0)\rangle
  +\ldots
  \enum
}
to hold at large $x_0$, up to exponentially decaying terms with
exponents larger than $M$,
$\drv{0}$ and $\drvstar{0}$ being the standard forward and backward
difference operators.
Insertion of eqs.~(C.2),(C.3) then leads
to the exact relation
\equation{
  m={M\ring{M}\imp{F}\over G}-\frac{1}{4}\cA\hat{M}^4,
  \qquad
  \ring{M}=\sinh(M),
  \qquad
  \hat{M}=2\sinh(M/2).
  \enum
}
The term proportional to $\cA$ in this equation
is a lattice artefact of order $a^3$, which could be removed by replacing the
forward and backward difference operators in eq.~(C.4) by
$\ring{\partial}_0$.

\subsection C.2 Renormalization and dependence on $\cA$

As explained in ref.~[\ref{QuarkFlow}],
the axial-current renormalization constant $\ZAh$ can be
computed by probing the PCAC relation
at positive gradient-flow time. The
renormalization constant calculated in this way includes
the $\rmO(am)$ corrections required for
$\rmO(a)$-improvement [\ref{SFimp},\ref{ImpNonD}].

The chiral Ward identities used to determine $\ZAh$
(eqs.~(8.8) and (8.10) in ref.~[\ref{QuarkFlow}])
are relations among two-point correlation functions summed over
a range $[-d,d]$ of time $x_0$. At large $d$ the calculated values
of $\ZAh$ very rapidly become independent of $d$, the leading
corrections decaying exponentially with
exponents equal to the next-to-lowest energies in the flavour channel
considered. The renormalization constant is then determined
in the range of $d$, where these corrections can be safely neglected.

The improvement coefficient $\cA$ appears in the Ward identities
implicitly, through the quark-mass sum $m$, and explicitly
multiplying one of the correlation functions.
Since the latter decays exponentially at large $d$,
and since $\ZAh$ and $m$ only occur in the
combination $\ZAh m$ in this limit, it follows that the product $\ZAh m$
is independent of $\cA$.
Recalling the PCAC relation (C.5), this implies that
the renormalized improved decay constant $\ZAh\imp{F}$
is insensitive to the value of $\cA$ up to
a tiny contribution of order $a^3$
deriving from the term proportional to $\cA$ in eq.~(C.5).

\subsection C.3 Table of renormalization factors

The results obtained in the runs \Ai--\Ci{}
for the quark-mass sums and the renormalization
constants in the $ud$ and $us$ flavour channels
are listed in table~5.
While $\Fpi$ and $\FK$ are practically independent of $\cA$,
the numbers quoted in the table are not and a choice of $\cA$ thus
had to be made (second column in table~5).

\topinsert
\newdimen\digitwidth
\setbox0=\hbox{\rm 0}
\digitwidth=\wd0
\catcode`@=\active
\def@{\kern\digitwidth}
\tablecaption{PCAC quark-mass sums and axial-current renormalization constants}
\vskip-3.5ex

$$\vbox{\settabs\+&%
                  xxxx&&%
                  xxxxxxxxxxxx&x&%
                  xxxxxxxxxxxx&x&%
                  xxxxxxxxxxxx&x&%
                  xxxxxxxxxxxx&x&%
                  xxxxxxxxxxxx&\cr
\thicktablerule
\vskip1.2ex
                \+& \hfill Run\hfill
                 && \hfill $\cA$\hfill
                 && \hfill $am_{ud}$\hfill
                 && \hfill $am_{us}$\hfill
                 && \hfill $\ZAh^{ud}$\hfill
		 && \hfill $\ZAh^{us}$\hfill
                 &\cr
\vskip0.8ex
\thintablerule
\vskip1.2ex
{\ninepoint
  \+& \hfill \Ai\hfill
  &&  \hfill $-0.039@$\hfill
  &&  \hfill $0.021072(81)$\hfill
  &&  \hfill $-$\hfill
  &&  \hfill $0.8075(11)$\hfill
  &&  \hfill $-$\hfill
&\cr
  \+& \hfill \Aii\hfill
  &&  \hfill $-0.039@$\hfill
  &&  \hfill $0.010581(77)$\hfill
  &&  \hfill $0.026315(65)$\hfill
  &&  \hfill $0.8054(24)$\hfill
  &&  \hfill $0.8079(10)$\hfill
  &\cr
  \+& \hfill \Aiii\hfill
  &&  \hfill $-0.039@$\hfill
  &&  \hfill $0.00556(14)@$\hfill
  &&  \hfill $0.028321(99)$\hfill
  &&  \hfill $0.8132(63)$\hfill
  &&  \hfill $0.8093(12)$\hfill
  &\cr
  \+& \hfill \Bi\hfill
  &&  \hfill $-0.035@$\hfill
  &&  \hfill $0.014502(41)$\hfill
  &&  \hfill $-$\hfill
  &&  \hfill $0.8138(7)@$\hfill
  &&  \hfill $-$\hfill
  &\cr
  \+& \hfill \Ci\hfill
  &&  \hfill $-0.0533$\hfill
  &&  \hfill $0.01924(17)@$\hfill
  &&  \hfill $-$\hfill
  &&  \hfill $0.8066(22)$\hfill
  &&  \hfill $-$\hfill
  &\cr
}
\vskip1.0ex
\thicktablerule
}
$$
\vskip0.0ex
\endinsert

In the case of run \Ci{}, where the traditional form of the Dirac
operator was used, $\cA$ was set to the non-perturbatively
determined value obtained in ref.~[\ref{CAThree}].
Lacking a similarly systematic determination of the coefficient
in the theory with the modified Dirac operator,
the other values of $\cA$ listed in the table
were estimated directly on the simulated lattices
using the so-called LANL method [\ref{LANL},\ref{LANLimp}].
To leading order, the perturbation expansion
\equation{
  \cA=-0.00603(3)\times g_0^2+\rmO(g_0^4),
  \enum
}
incidentally coincides with the one in the theory with the
traditional form of the Dirac operator and tree-level improved
gauge action [\ref{cAPThI},\ref{cAPThII}],
since the relevant axial-current vertex diagram is the same.
The coefficients of the
$\rmO(am)$ contributions to the axial-current renormalization constant
are, to this order,
unchanged as well
and the remark applies in the case of
other composite fields too.

On coarse lattices, the calculated
values of $\cA$ may depend quite a bit
on the chosen improvement condition. Varying $\cA$ by
$\pm0.05$ however affects the renormalized decay constants only
at a level of a quarter of a per mille, i.e.~by amounts roughly an order
of magnitude smaller than the statistical errors quoted in table~3.

\appendix D. Eigenvalue distributions

The low-lying eigenvalues of $(\DD)^{1/2}$ (where $D$ is
the light-quark lattice Dirac operator) were computed to
a relative precision of $0.5\%$ using the Chebyshev-accelerated
subspace iteration described in appendix A of ref.~[\ref{Stability}].
This method delivers both the eigenvalues and eigenvectors,
which allows the accuracy of the eigenvalues to be rigorously
controlled.

\topinsert
\vbox{
\vskip0ex
\epsfxsize=11.4cm\hskip0.35cm%
\epsfbox{plots/evahist.eps}%
\vskip1.0ex
\figurecaption{%
Normalized distributions of the lowest eigenvalue $\lambda$ of
$(\DD)^{1/2}$. The dotted lines indicate the position of
the median of the distributions and the arrows the one of
$\frac{1}{2}\ZAh^{ud}m_{ud}$ (cf.~table~4).
}
}
\vskip0.0ex
\endinsert

Some of the simulations reported in sect.~6 use twisted-mass reweighting
for the light quarks [\ref{TMopenQCD}]
and all use a rational approximation for
the pseudo-fermion representation of the strange-quark determinant
[\ref{RHMCI},\ref{RHMCII}].
The probability densities $\rho(\lambda)$
of the lowest eigenvalue $\lambda$ of $(\DD)^{1/2}$
shown in fig.~6 take the associated reweighting factors into account
and so do the medians $\bar{\mu}$ and widths $\sigma$ of
the distributions quoted in table~4.

More precisely, the latter are defined as follows. Consider
an ensemble of $N$ gauge fields with normalized
weights $w_1,\ldots,w_N$ and let
$\lambda_1,\ldots,\lambda_N$ be the computed lowest eigenvalues
of $(\DD)^{1/2}$. The empirical probability for the eigenvalue
to be less than or equal to $\lambda$ is given by
\equation{
  P_N(\lambda)=\sum_{k=1}^Nw_k\left(\lambda_k\leq\lambda\right),
  \enum
}
where the bracket is $1$ if the enclosed condition is true and $0$ otherwise.
$P_N(\lambda)$ is a step function that increases monotonically from $0$ to $1$.
The median of the eigenvalue distribution is then equal to $(u+v)/2$,
$u\leq v$ being the largest and smallest step points satisfying
$P_N(u)\leq0.5$ and $P_N(v)\geq0.5$, respectively.
And its width is {\it half}\/ the size $|v-u|$ of the smallest
interval $[u,v]$ such that
$\sum_{k=1}^Nw_k\left(u\leq\lambda_k\leq v\right)\geq0.683$.

\beginbibliography


\bibitem{LuscherMaster}
M. L\"uscher,
{\it Stochastic locality and master-field simulations of very large lattices},
EPJ Web Conf. 175 (2018) 01002

\bibitem{HighTop}
L. Giusti, M. L\"uscher,
{\it Topological susceptibility at
$T>T_{\rm c}$ from master-field simulations of the SU(3) gauge theory},
Eur. Phys. J. C79 (2019) 207


\bibitem{Wilson}
K. G. Wilson, {\it Confinement of quarks}, Phys. Rev. D10 (1974) 2445


\bibitem{SW}
B. Sheikholeslami, R. Wohlert,
{\it Improved continuum limit lattice action for QCD with Wilson fermions},
Nucl. Phys. B259 (1985) 572

\bibitem{SFimp}
M. L\"uscher, S. Sint, R. Sommer, P. Weisz,
{\it Chiral symmetry and O(a) improvement in lattice QCD},
Nucl. Phys. B478 (1996) 365


\bibitem{Horowitz}
A. M. Horowitz,
{\it Stochastic quantization in phase space},
Phys. Lett. 156B (1985) 89;
{\it The second order Langevin equation and numerical simulations},
Nucl. Phys. B280 [FS18] (1987) 510;
{\it A generalized guided Monte Carlo algorithm},
Phys. Lett. B268 (1991) 247

\bibitem{JansenLiu}
K. Jansen, C. Liu,
{\it
Kramers equation algorithm for simulations of QCD with two flavors of
Wilson fermions and gauge group SU(2)},
Nucl.Phys. B453 (1995) 375 [E: {\it ibid.} B459 (1996) 437]


\bibitem{HMC}
S. Duane, A. D. Kennedy, B. J. Pendleton, D. Roweth,
{\it Hybrid Monte Carlo},
Phys. Lett. B195 (1987) 216


\bibitem{SF}
M. L\"uscher, R. Narayanan, P. Weisz, U. Wolff,
{\it The Schr\"odinger functional: A renormalizable probe for
non-Abelian gauge theories},
Nucl. Phys. B384 (1992) 168

\bibitem{SFQ}
S. Sint,
{\it On the Schr\"odinger functional in QCD},
Nucl. Phys. B421 (1994) 135


\bibitem{OpenBC}
M. L\"uscher, S. Schaefer,
{\it Lattice QCD without topology barriers},
JHEP 1107 (2011) 036


\bibitem{TMopenQCD}
M. L\"uscher, S. Schaefer,
{\it
Lattice QCD with open boundary conditions and twisted-mass reweighting},
Comput. Phys. Commun. 184 (2013) 519


\bibitem{TreeSymanzik}
P. Weisz,
{\it Continuum limit improved lattice action for pure Yang-Mills theory (I)},
Nucl. Phys. B212 (1983) 1


\bibitem{CswThree}
J. Bulava, S. Schaefer,
{\it Improvement of $N_f=3$ lattice QCD with Wilson\hfill\break
fermions and tree-level improved gauge action},
Nucl. Phys. B874 (2013) 188


\bibitem{Ergodicity}
M. L\"uscher,
{\it Ergodicity of the SMD algorithm in lattice QCD},
unpublished notes (2017),
{\tt http://cern.ch/luscher/notes/smd-ergodicity.pdf}


\bibitem{OMF}
I. P. Omelyan, I. M. Mryglod, R. Folk,
{\it Symplectic analytically integrable decomposition
algorithms: classification, derivation, and application to molecular
dynamics, quantum and celestial mechanics simulations},
Comp. Phys. Commun. 151 (2003) 272


\bibitem{chrono}
R. C. Brower, T. Ivanenko, A. R. Levi, K. N. Orginos,
{\it Chronological inversion method for the Dirac
matrix in hybrid Monte Carlo},
Nucl. Phys. B484 (1997) 353


\bibitem{DFL}
M. L\"uscher,
{\it Local coherence and deflation of the low quark modes in lattice QCD},
JHEP 0707 (2007) 081;
{\it Deflation acceleration of lattice QCD simulations},
JHEP 0712 (2007) 011


\bibitem{Knuth}
D. E. Knuth,
{\it Semi-Numerical Algorithms}, {\it in\/}:
The Art of Computer Programming, vol. 2, 2nd ed.
(Addison-Wesley, Reading MA, 1981)


\bibitem{Urbach}
C. Urbach,
{\it Reversibility violation in the Hybrid Monte Carlo algorithm},
Comput. Phys. Commun. 224 (2018) 44


\bibitem{Dekker}
T. J. Dekker,
{\it A floating-point technique for extending the available precision},
Numer. Math. 18 (1971) 224

\bibitem{Shewchuk}
J. R. Shewchuk, {\it Adaptive precision floating-point arithmetic
and fast robust geo\-metric predicates},
Discrete \& Computational Geometry 18 (1997) 305


\bibitem{Saad}
Y. Saad,
{\it Iterative methods for sparse linear systems}, 2nd ed.
(SIAM, Philadelphia, 2003); see also
{\tt http://www-users.cs.umn.edu/\~{}saad/}.


\bibitem{Hasenbusch}
M. Hasenbusch,
{\it Speeding up the Hybrid Monte Carlo algorithm for dynamical fermions},
Phys. Lett. B519 (2001) 177

\bibitem{HasenbuschJansen}
M. Hasenbusch, K. Jansen,
{\it Speeding up lattice QCD simulations with clover-improved Wilson
fermions}, Nucl. Phys. B659 (2003) 299

\bibitem{DDHMC}
M. L\"uscher,
{\it Schwarz-preconditioned HMC algorithm for two-flavor lattice QCD},
Comp. Phys. Commun. 165 (2005) 199

\bibitem{UrbachEtAl}
C. Urbach, K. Jansen, A. Shindler, U. Wenger,
{\it HMC algorithm with multiple time scale
integration and mass preconditioning},
Comp. Phys. Commun. 174 (2006) 87

\bibitem{RHMCI}
I. Horv\'ath, A. D. Kennedy, S. Sint,
{\it A new exact method for dynamical fermion computations with
non-local actions},
Nucl. Phys. (Proc. Suppl.) 73 (1999) 834

\bibitem{RHMCII}
M. A. Clark, A. D. Kennedy,
{\it Accelerating dynamical fermion computations using the
Rational Hybrid Monte Carlo (RHMC) algorithm with
multiple pseudo-fermion fields},
Phys. Rev. Lett. 98 (2007) 051601


\bibitem{cswI}
M. L\"uscher, S. Sint, R. Sommer, P. Weisz, U. Wolff,
{Nonperturbative $\rmO(a)$ improvement of lattice QCD},
Nucl.Phys. B491 (1997) 323-343


\bibitem{CLSI}
M. Bruno et al.,
{\it Simulation of QCD with $N_f=2+1$
flavors of non-perturbatively improved Wilson fermions},
JHEP 1502 (2015) 043

\bibitem{CLSII}
M. Bruno, T. Korzec, S. Schaefer,
{\it Setting the scale for the CLS 2+1 flavor ensembles},
Phys. Rev. D95 (2017) 074504


\bibitem{WilsonFlow}
M. L\"uscher,
{\it Properties and uses of the Wilson flow in lattice QCD},
JHEP 1008 (2010) 071 [Erratum: {\it ibid.} 1403 (2014) 092]



\bibitem{QuarkFlow}
M. L\"uscher,
{\it Chiral symmetry and the Yang--Mills gradient flow},
JHEP 1304 (2013) 123


\bibitem{ImpNonD}
T. Bhattacharya, R. Gupta, W. Lee, S. R. Sharpe, J. M. S. Wu,
{\it Improved bilinears in lattice QCD with non-degenerate quarks},
Phys. Rev. D73 (2006) 034504


\bibitem{OQCD}
{\tt http://cern.ch/luscher/openQCD}


\bibitem{Stability}
L. Del Debbio, L. Giusti, M. L\"uscher, R. Petronzio, N. Tantalo,
{\it Stability of lattice QCD simulations and the thermodynamic limit},
JHEP 0602 (2006) 011

\bibitem{StabilityII}
L. Del Debbio, L. Giusti, M. L\"uscher, R. Petronzio, N. Tantalo,
{\it QCD with light Wilson quarks on fine lattices (II):
DD-HMC simulations and data analysis},
JHEP 0702 (2007) 082


\bibitem{XSF}
M. Dalla Brida, T. Korzec, S. Sint, P. Vilaseca,
{\it High precision renormalization of the flavour non-singlet
Noether currents in lattice QCD with Wilson quarks},
Eur. Phys. J. C79 (2019) 23


\bibitem{ZAThree}
J. Bulava, M. Della Morte, J. Heitger, C. Wittemeier,
{\it Nonperturbative renormalization of the axial current in $N_f=3$
lattice QCD with Wilson fermions and tree-level improved gauge action},
Phys. Rev. D93 (2016) 114513


\bibitem{CAThree}
J. Bulava, M. Della Morte, J. Heitger, C. Wittemeier,
{\it Non-perturbative improvement of the axial current in $N_f=3$
lattice QCD with Wilson fermions and tree-level improved gauge action},
Nucl. Phys. B896 (2015) 555


\bibitem{LANL}
T. Bhattacharya, R. Gupta, W. Lee, S. Sharpe,
{\it Order ``a'' improved renormalization constants},
Phys. Rev. D63 (2001) 074505

\bibitem{LANLimp}
S. Collins, C. T. H. Davies, G. P. Lepage, J. Shigemitsu,
{\it A nonperturbative
determination of the O(a) improvement coefficient $\cA$
and the scaling of $f_{\pi}$ and $m^{\MSbar}$},
Phys. Rev. D67 (2003) 014504


\bibitem{cAPThI}
Y. Taniguchi, A. Ukawa,
{\it Perturbative calculation of improvement
coefficients to $\rmO(g^2a)$ for bilinear quark operators in lattice QCD},
Phys. Rev. D58 (1998) 114503

\bibitem{cAPThII}
S. Aoki, R. Frezzotti, P. Weisz,
{\it Computation of the improvement
coefficient $\csw$ to 1-loop with improved gluon actions},
Nucl. Phys. B540 (1999) 501

\endbibliography

\bye